\documentclass[aip%
 aip,
 amsmath,amssymb,
 reprint,%
]{revtex4-1}

\usepackage{graphicx}
\usepackage{dcolumn}
\usepackage{bm}

\usepackage[utf8]{inputenc}
\usepackage[T1]{fontenc}
\usepackage{mathptmx}
\usepackage{etoolbox}
\usepackage{float}
\usepackage{booktabs}
\usepackage{subcaption}
\usepackage{soul}
\usepackage[normalem]{ulem}
\usepackage[most]{tcolorbox}
\newtcolorbox{promptblock}{
  colback=gray!5,
  colframe=black!60,
  boxrule=0.5pt,
  arc=2pt,
  left=6pt,
  right=6pt,
  top=6pt,
  bottom=6pt,
  fonttitle=\bfseries,
  title=Prompt
}

\usepackage[
    colorlinks=true,
    linkcolor=blue,
    citecolor=blue,
    urlcolor=blue
]{hyperref}

\makeatletter
\def\@email#1#2{%
 \endgroup
 \patchcmd{\titleblock@produce}
  {\frontmatter@RRAPformat}
  {\frontmatter@RRAPformat{\produce@RRAP{*#1\href{mailto:#2}{#2}}}\frontmatter@RRAPformat}
  {}{}
}%
\makeatother
\begin{document}

\preprint{AIP/123-QED}

\title[Chaos in reason]{Chaos in reason:\\How chain-of-thought LLMs can look for an answer}

\author{Gregorio Jaca}
\affiliation{Department of Theoretical Physics, Institute of Physics, Budapest University of Technology and Economics, Műegyetem rkp. 3, Budapest, H-1111, Hungary}
\author{János Török}%
\affiliation{Department of Theoretical Physics, Institute of Physics, Budapest University of Technology and Economics, Műegyetem rkp. 3, Budapest, H-1111, Hungary}
\affiliation{HUN-REN-BME Morphodynamics Research Group, Műegyetem rkp. 3, Budapest, 1111, Hungary}
\email{torok.janos@ttk.bme.hu}

\author{Kristóf Benedek}
\affiliation{Department of Theoretical Physics, Institute of Physics, Budapest University of Technology and Economics, Műegyetem rkp. 3, Budapest, H-1111, Hungary}
\affiliation{Institute of Technical Physics and Materials Science, HUN-REN Centre for Energy Research, Konkoly-Thege Miklós út 29-33, P.O. Box 49, Budapest, H-1525, Hungary}%

\date{\today}

\begin{abstract}
Large Language Models (LLMs) have achieved remarkable performance across a wide range of tasks, yet their internal dynamics remain poorly understood. In this work, we apply the tools of nonlinear dynamics and chaos theory to LLMs. By analyzing both text and hidden state trajectories, we demonstrate that LLMs exhibit hallmark signatures of chaos, including strong sensitivity to initial conditions, manifested as intermittent, jump-like divergence of nearby trajectories combined with bounded evolution, with consistent results across different distance metrics. An exact Jacobian analysis of the Transformer's sub-blocks shows that self-attention and the feed-forward network expand and propagate perturbations, while normalization and residual connections counteract this expansion and promote stability. Recurrence plots show structural similarities between LLMs and canonical chaotic systems such as the Lorenz attractor, while dimension analysis reveals fractal structures in the hidden state space, particularly pronounced in the last layers. We propose that the nonlinear coupling induced by attention mechanisms plays a key role in driving this chaotic behavior.
\end{abstract}

\maketitle

\section{Introduction: How does IT think?}\label{sec:intro}
Since the introduction of the transformer architecture with Attention Is All You Need in 2017 \cite{attention}, large language models (LLMs) have rapidly evolved into highly capable systems, culminating in their widespread adoption and public impact beginning in 2022. These models exhibit striking emergent abilities, particularly in reasoning-intensive tasks. A notable milestone was the observation that prompting models to generate intermediate reasoning steps-so-called chain-of-thought (CoT) reasoning-substantially improves performance on complex problems \cite{wei2023chainofthoughtpromptingelicitsreasoning}. This discovery raised fundamental questions about the nature of reasoning in artificial systems and the mechanisms by which such behavior arises.

In human cognition, reasoning is traditionally associated with logical inference, a formal discipline with thousands of years old deep roots in philosophy and mathematics. By contrast, LLMs are trained through large-scale statistical optimization and operate without explicit symbolic rules. This contrast invites a natural question: How is this reasoning perceived by a machine, in particular by an LLM? Does it follow some pattern, or is it stochastic? Does it have a simple or a complex structure? Or is it by any chance chaotic?

From a theoretical standpoint, LLMs can be viewed as high-dimensional nonlinear dynamical systems. At each generation step, the model updates an internal hidden state via a sequence of nonlinear transformations involving self-attention and feedforward layers. The autoregressive generation of tokens corresponds to a discrete-time evolution rule, with the initial prompt acting as an initial condition. Under this interpretation, a generated text sequence after some layers, for instance, in the latent space, can be viewed as a trajectory.

This perspective enables a direct analogy between classical chaotic systems and LLMs, summarized in Table \ref{tab:analogy}. Concepts such as phase space, trajectories, nonlinear coupling, and sensitivity to initial conditions have natural counterparts in the context of machine learning. In particular, the self-attention mechanism introduces strong nonlinear coupling between components of the hidden state, a structural feature reminiscent of the stretching-and-folding mechanisms that underlie chaos in low-dimensional systems.

\begin{table}[h]
\centering
\begin{tabular}{@{}ll@{}}
\toprule
\textbf{Chaotic System} & \textbf{LLMs} \\ \midrule
Phase Space & Embedding / Hidden State Space \\
State Vector & Hidden State Vector \\
Phase-Space Trajectory & Sequence of Token Text or Hidden States \\
Continuous Time Evolution & Discrete Token Generation Step \\
Differential Equation & Autoregressive Function \\
Numerical Integration & Forward-Pass Inference \\
Nonlinear Coupling & Self-Attention Mechanism \\
Stretching & Attention / MLP Layers \\
Folding & Normalization Layers \\
Initial Condition & Initial Prompt Text or Embedding \\
Measurement Uncertainty & Floating-Point Precision, Quantization \\ \bottomrule
\end{tabular}
\caption{Analogies between concepts in chaos theory and LLMs.}
\label{tab:analogy}
\end{table}

Modern LLMs are known to exhibit stable generative behavior: they produce fluent, contextually consistent text over long sequences, while simultaneously allowing for variability in their outputs. This coexistence of boundedness and sensitivity is a defining characteristic of chaotic systems in nonlinear dynamics. Chaotic systems evolve within a constrained region of phase space yet display exponential sensitivity to perturbations in their initial conditions, quantified by positive Lyapunov exponents. The similarity is suggestive: small changes in prompts, embeddings, or numerical precision can lead to qualitatively different generation steps, reasoning steps, despite the overall coherence of the model’s behavior.

Understanding whether the internal evolution of LLM representations exhibits genuine signatures of chaos is therefore of both theoretical and practical interest. From a scientific perspective, it offers a principled framework for interpreting model behavior beyond purely statistical descriptions. From an engineering standpoint, it may shed light on issues such as reproducibility, robustness, and controllability in large-scale generative models.

In this work, we apply tools from nonlinear dynamics and chaos theory to the analysis of LLMs. By examining both token-level outputs and hidden-state trajectories, we investigate sensitivity to initial conditions, Lyapunov spectra, recurrence structures, and effective dimensionality of the latent space. Our results suggest that key hallmarks of chaotic dynamics are present in LLMs, with particularly rich structure emerging in the deeper layers of the network. We further argue that the nonlinear coupling induced by self-attention plays a central role in shaping this behavior. Beyond these global, trajectory-level diagnostics, we complement the analysis with an exact Jacobian characterization of the individual Transformer sub-blocks, self-attention, the feed-forward network, normalization, and the residual connection, in order to identify, at the level of single operators, which components amplify perturbations (stretching) and which suppress them (folding). This bridges the macroscopic chaotic signatures observed in generated trajectories with a microscopic, mechanistic picture of how each architectural component contributes to them.

\section{Previous work}\label{sec:prev_work}

The application of dynamical systems theory to modern neural networks is a relatively unexplored field, but might shed light to questions regarding interpretability and reproducibility. While early work primarily examined low-dimensional recurrent networks, recent studies have begun to analyze Transformer architectures and large language models (LLMs) through the lens of nonlinear dynamics, providing the foundation for the present work.

\paragraph{Chaotic Dynamics in Transformers and LLMs.}
Some recent studies explicitly characterize Transformers as nonlinear dynamical systems. Li \textit{et al.}~\cite{li2025cognitive_activation} argue that reasoning in LLMs arises from chaotic information extraction processes and introduce quasi-Lyapunov exponents to quantify layerwise sensitivity to perturbations. Geshkovski \textit{et al.}~\cite{geshkovski2025mathematicalperspectivetransformers} formalize self-attention as a nonlinear coupling mechanism between token representations, establishing a mathematical basis for dynamical interpretations of attention.

Analytically tractable models further support this view. Dynamical mean-field analyses of simplified self-attention networks~\cite{dynamicalmeanfieldtheoryselfattention} show that even binary-weight Transformers can exhibit non-equilibrium phase transitions and chaotic bifurcations, suggesting that such dynamics arise from architectural structure rather than scale alone. Complementary work by Tomihari \textit{et al.}~\cite{tomihari2025recurrent_self_attention_dynamics} demonstrates that normalization layers regulate the Jacobian spectrum, constraining instability and driving models toward a critical regime with maximum Lyapunov exponents close to zero, suggesting operation \textit{near the edge of chaos: a critical regime where signals neither explode nor vanish, enabling long-range information propagation}.

\paragraph{Jacobian and Spectral Analysis of Transformer Stability.}
A complementary line of work probes stability directly through the Jacobian of individual network components rather than through end-to-end trajectories. The classical starting point is the demonstration of a transition to chaos in random recurrent networks~\cite{sompolinsky1988chaos}, together with the observation that computational capability of recurrent networks peaks near the order--chaos boundary~\cite{bertschinger2004edge}. This tradition was developed further in the mean-field theory of randomly initialized feedforward networks, where signal and gradient propagation across depth were shown to undergo an order-to-chaos transition controlled by the spectrum of the layer Jacobian, with trainable networks confined to a critical ``edge of chaos'' manifold~\cite{poole2016exponential_expressivity,schoenholz2017deep_information_propagation}. Residual connections were later shown to soften this transition, turning exponential signal decay or blow-up into sub-exponential, polynomial dynamics~\cite{yang2017meanfield_resnets_edge_of_chaos}, and gated recurrent architectures were analyzed through the same lens~\cite{chen2018dynamical_isometry_rnn}. For Transformers specifically, the self-attention Jacobian has been shown to be state-dependent and not globally Lipschitz over an unbounded input domain~\cite{kim2021lipschitzconstantselfattention}, while the softmax nonlinearity it embeds admits an exact, dimension-independent Lipschitz bound of $1/2$~\cite{nair2025softmax12lipschitztightbound}, recently generalized into a full geometric characterization of attention sensitivity~\cite{emadi2026exactattentionsensitivitygeometry}. Collapsing attention entropy, i.e.\ attention rows saturating to near one-hot vectors, has independently been identified as a source of training instability, motivating explicit spectral reparametrizations or preconditioning of the attention Jacobian~\cite{zhai2023attention_entropy_collapse,saratchandran2026spectral_conditioning_attention}. At the scale of full production models, direct Jacobian eigendecomposition of the residual stream reveals a systematic transition from non-normal, rotation-dominated early layers to near-symmetric late layers, together with a low-rank bottleneck that funnels perturbations into a shrinking effective subspace with depth~\cite{fernando2026transformer_residual_stream_dynamics}, a trend consistent with the layer-dependent expansion and dimensionality reduction we report in Secs.~\ref{subsec:layer_perturbations} and~\ref{subsec:dimensionality}. Our own operator-level Jacobian analysis (Sec.~\ref{subsec:computational_jacobian}) is closest in spirit to this line of work, but is aimed at explaining the coexistence of stretching and folding required for chaos in a trained, autoregressive LLM, rather than at trainability or optimizer design.

\paragraph{Sensitivity, Determinism, and Attractors in Generative Models.}
Recent studies also focus on deterministic inference. He \textit{et al.}~\cite{he2025nondeterminism} show that GPU-level batch variance and floating-point effects can lead to divergent outputs, highlighting the intrinsic sensitivity of autoregressive generation. At a higher level, attractor-like behavior has been observed in generative models. Wang \textit{et al.}~\cite{wang2025unveiling_attractor_cycles} report convergence to periodic attractor cycles in LLM-based paraphrasing, leading to reduced linguistic diversity. Similar phenomena appear in image generation: analyses of CycleGAN models~\cite{cyclegan} reveal chaotic attractors with positive Lyapunov exponents and effective dimensions comparable to the intrinsic dimensionality of the training data. These results illustrate how chaotic dynamics can coexist with bounded generation on a learned manifold.

\paragraph{Dynamical Perspectives on Interpretability.}

Dynamical systems tools have also been used to interpret neural computations. Sussillo and Barak~\cite{sussillo2013} showed that fixed points and local linearization in trained recurrent networks provide interpretable descriptions of network function. More recently, Zhang \textit{et al.}~\cite{zhang2024intelligence_edge_of_chaos} argue that optimal performance emerges at at an optimal complexity level, a regime neither fully ordered nor chaotic or random. Systems with overly ordered or overly chaotic regimes exhibit degraded behavior. In the context of language models, Zhou \textit{et al.}~\cite{zhou2025geometryreasoningflowinglogics} interpret chain-of-thought reasoning as smooth flows in embedding space, aligning naturally with a dynamical view of inference. This dynamical perspective complements the mechanistic-interpretability programme, which explains model behavior bottom-up by identifying circuits and analyzing attention patterns~\cite{olah2020zoom,ameisen2025circuit,lindsey2025biology}; chaos-theoretic diagnostics instead quantify emergent global properties such as stability, divergence, and attractor structure.

\paragraph{Scope of the Present Work.}

While prior studies have identified individual dynamical signatures in Transformers and generative models, a unified chaos-theoretic analysis of LLMs remains limited. The present work complements existing approaches by systematically analyzing both hidden-state trajectories and generated text using multiple diagnostics from nonlinear dynamics, enabling a cohesive characterization of chaotic behavior in large language models. Crucially, we connect the two traditions surveyed above explicitly: alongside trajectory-level diagnostics (divergence, recurrence, dimensionality), we derive and numerically measure the exact Jacobians of each architectural sub-block, self-attention, the feed-forward network, normalization, and the residual connection, in a trained, autoregressive LLM. This provides a component-resolved, mechanistic account of which operators stretch and which fold, directly linking the mean-field/Jacobian-spectrum tradition of Sec.~\ref{sec:prev_work} with the trajectory-based chaos diagnostics used throughout the rest of this paper.

\section{Preliminaries}
\label{sec:preliminaries}

This section introduces the concepts from nonlinear dynamics and chaos theory required to interpret large language models (LLMs) as dynamical systems. We then describe the transformer architecture from a latent-state perspective and formalize autoregressive inference as a discrete-time evolution in high-dimensional space. These preliminaries provide the conceptual and mathematical foundation for the methods introduced in Sec.~\ref{sec:methods}.

\subsection{Nonlinear Dynamics and Chaos}
\label{subsec:nonlinear_dynamics}

A \textbf{dynamical system} describes the evolution of a system’s state over time according to deterministic rules. The state is represented by a vector $\mathbf{x}(t)$ in a \textbf{phase space}, whose coordinates correspond to the system’s degrees of freedom. In continuous time, the evolution is governed by ordinary differential equations,
\begin{equation}
\frac{d\mathbf{x}}{dt} = \mathbf{F}(\mathbf{x}),
\end{equation}
while in discrete time it is described by iterated maps,
\begin{equation}
\mathbf{x}_{t+1} = \mathbf{f}(\mathbf{x}_t).
\end{equation}
A system is \textbf{deterministic} if its future evolution is uniquely determined by its current state. The resulting trajectory traces the system’s evolution in phase space, and its geometry encodes the qualitative behavior of the dynamics.

Chaotic systems form a subclass of deterministic nonlinear systems characterized by aperiodic behavior and \textbf{sensitivity to initial conditions}. Two trajectories starting from infinitesimally close initial states, $\mathbf{x}(0)$ and $\mathbf{x}(0)+\boldsymbol{\delta}(0)$, diverge exponentially in time,
\begin{equation}
\|\boldsymbol{\delta}(t)\| \approx \|\boldsymbol{\delta}(0)\| e^{\lambda t},
\end{equation}
where $\lambda$ is the \textbf{maximal Lyapunov exponent}. A positive maximal Lyapunov exponent is a defining signature of chaos, indicating exponential amplification of arbitrarily small uncertainties.

\begin{figure}[H]
    \centering
    \begin{subfigure}{0.48\linewidth}
        \centering
        \includegraphics[width=\linewidth]{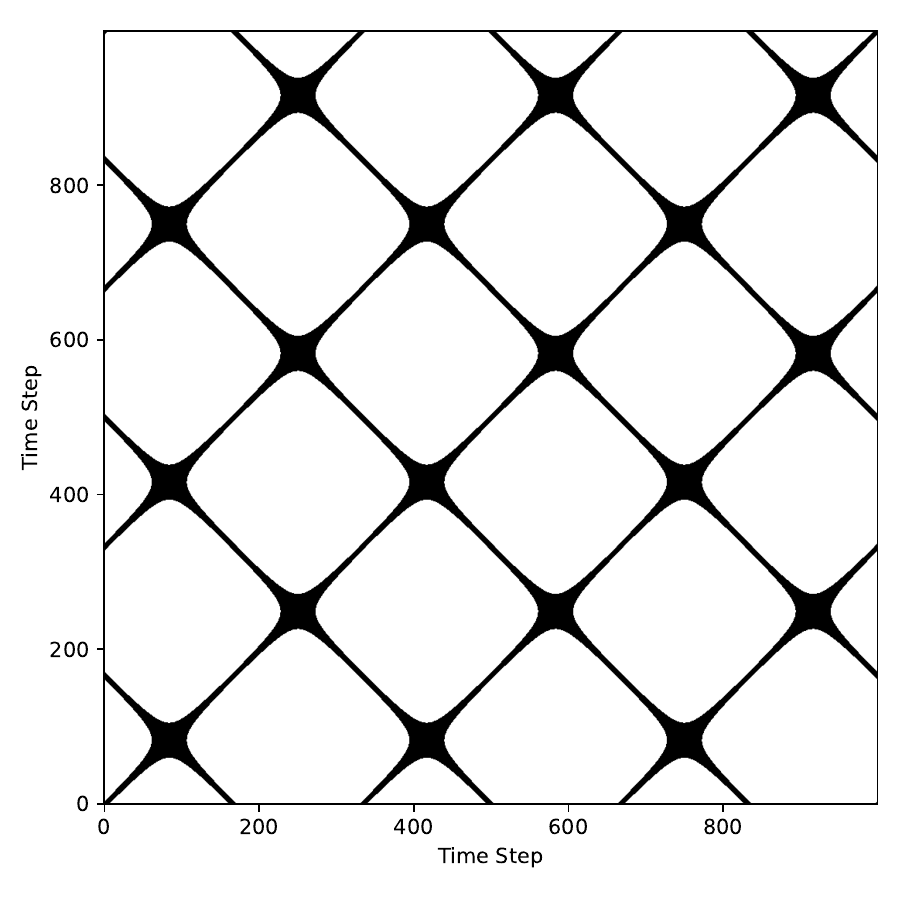}
        \caption{}
    \end{subfigure}
    \hfill
    \begin{subfigure}{0.48\linewidth}
        \centering
        \includegraphics[width=\linewidth]{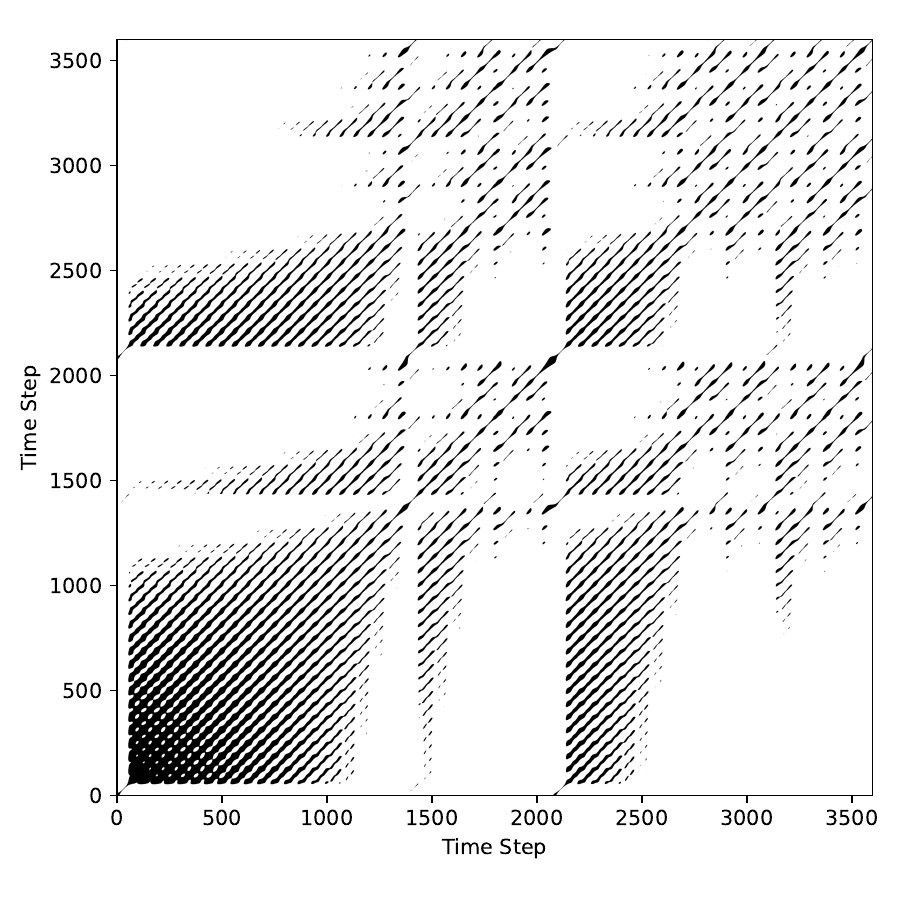}
        \caption{}
    \end{subfigure}

    \vspace{2pt} 

    \begin{subfigure}{0.48\linewidth}
        \centering
        \includegraphics[width=\linewidth]{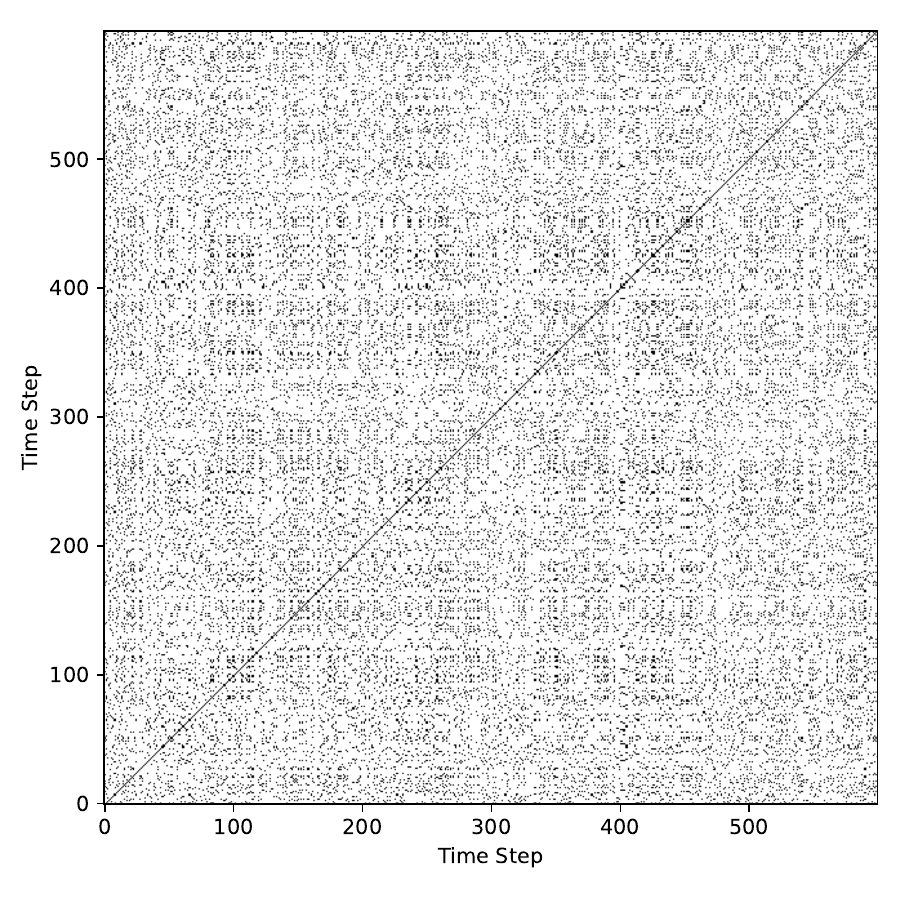}
        \caption{}
    \end{subfigure}
    \hfill
    \begin{subfigure}{0.48\linewidth}
        \centering
        \includegraphics[width=\linewidth]{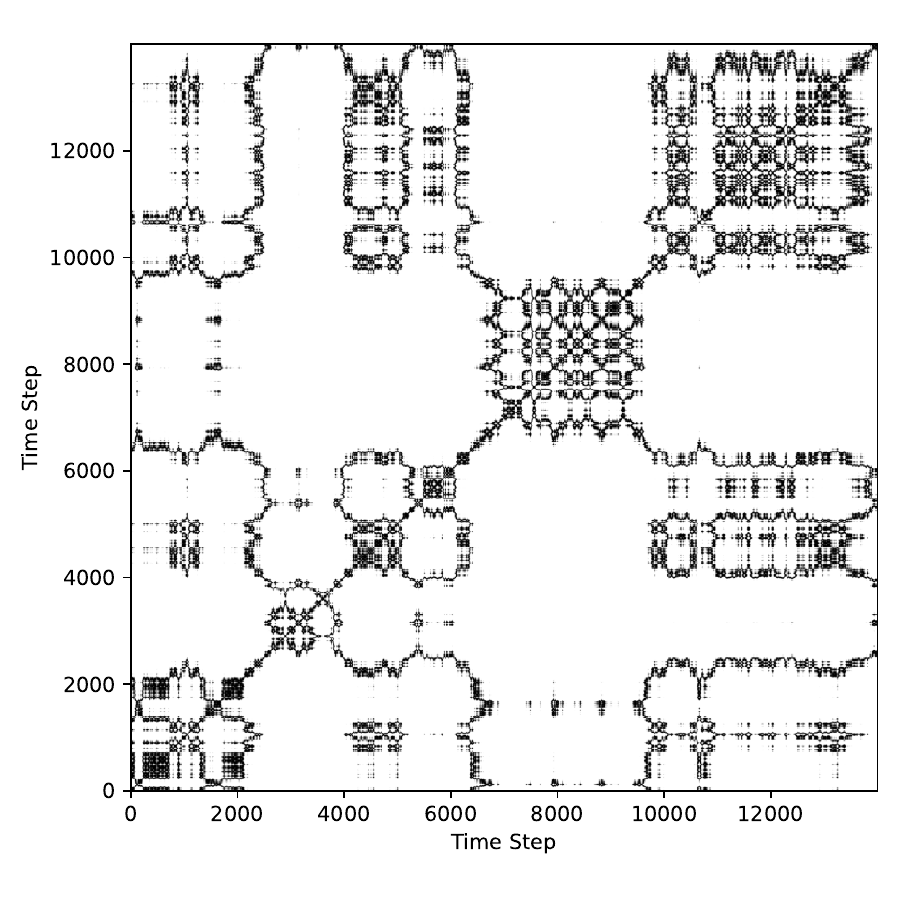}
        \caption{}
    \end{subfigure}

    \caption{Exemplary recurrence plots of (a) a periodic motion with one frequency, (b) the chaotic Lorenz system, (c) of normally distributed white noise, (d) brownian motion.}
    \label{fig:rp_examples}
\end{figure}

In an $n$-dimensional phase space, a dynamical system admits a Lyapunov spectrum $\{\lambda_1,\dots,\lambda_n\}$, characterizing expansion and contraction along different directions. Chaotic systems simultaneously exhibit stretching ($\lambda_i>0$) and contraction ($\lambda_i<0$), leading to the formation of \textbf{strange attractors}: bounded invariant sets with fractal geometry and non-integer effective dimensionality.

A commonly used measure of attractor geometry is the \textbf{correlation dimension} $D_2$.
Periodic systems yield integer values, while chaotic attractors exhibit non-integer dimensions reflecting fractal structure.


\subsection{Large Language Models as Dynamical Systems}
\label{subsec:llms_dynamics}

Large language models are deep neural networks based on the Transformer architecture, trained to model conditional token distributions and used autoregressively during inference. From a latent-space perspective, this autoregressive process naturally defines a discrete-time dynamical system.

Let the model vocabulary be
\begin{equation}
\mathcal{V} = \{1,\dots,|\mathcal{V}|\},
\end{equation}
and let the learned token embedding map be
\begin{equation}
\Phi : \mathcal{V} \rightarrow \mathbb{R}^D,
\end{equation}
where $D$ is the hidden dimension. A prompt of length $L_p$ with tokens $(v_1,\dots,v_{L_p})$ is mapped to an initial embedding sequence
\begin{equation}
\mathbf{E}_0 = (\mathbf{e}_1,\dots,\mathbf{e}_{L_p}), 
\qquad 
\mathbf{e}_i = \Phi(v_i).
\end{equation}

During inference, the transformer evolves a context window of hidden states of size $L$. Let
\begin{equation}
\mathbf{H}_t = (\mathbf{h}_{t-L+1},\dots,\mathbf{h}_t) \in \mathbb{R}^{L \times D}
\end{equation}
denote the hidden states available at generation step $t$. The forward pass defines a deterministic update rule,
\begin{equation}
\mathbf{H}_{t+1} = F(\mathbf{H}_t),
\end{equation}
where $F$ represents the composition of self-attention, feedforward, and normalization layers.

At each step, the context window is updated by discarding the oldest state and appending the newly generated one, inducing a nonlinear discrete-time mapping on a bounded subset of $\mathbb{R}^{L \times D}$. For analysis, we focus on the hidden state of the most recent token,
\begin{equation}
\mathbf{h}_t \in \mathbb{R}^D,
\end{equation}
and treat the sequence
\begin{equation}
\{\mathbf{h}_t\}_{t=1}^T\equiv (h_1, h_2,\ldots,h_T)
\end{equation}
as a \textbf{trajectory} in latent space.

Two complementary trajectory notions arise naturally in this setting. A \textbf{horizontal} trajectory follows the hidden state of the newest token across generation steps $t$ at a fixed layer; it is the object of the divergence, recurrence, and dimensionality analyses below. A \textbf{vertical} trajectory follows the hidden state of a single token across the layers $\ell = 1\dots L$ within one forward pass; since every layer has its own weights, this map is non-autonomous, and it is the object of the operator-level perturbation analysis of Sec.~\ref{subsec:layer_perturbations}.

Self-attention introduces nonlinear, state-dependent coupling between token representations within the context window, enabling amplification of small perturbations. Feedforward (MLP) layers further contribute to nonlinear transformations, while normalization layers constrain activation magnitudes, ensuring bounded trajectories. This alternation of stretching and folding operations parallels the geometric mechanisms underlying classical chaotic systems.

\subsection{Initial Conditions and Perturbations}
\label{subsec:init_conditions}

To probe sensitivity to initial conditions, we treat small perturbations of the initial embedding tensor $\mathbf{E}_0$ as nearby initial states of the dynamical system. Given a fixed prompt embedding $\mathbf{E}_0 \in \mathbb{R}^{L \times D}$, we generate perturbed initial conditions
\begin{equation}
\mathbf{E}_0^{(k)} = \mathbf{E}_0 + \boldsymbol{\Delta}^{(k)},
\end{equation}
where $\boldsymbol{\Delta}^{(k)} \in \mathbb{R}^{L \times D}$ is a random perturbation of fixed norm,
\begin{equation}
\|\boldsymbol{\Delta}^{(k)}\| = r.
\end{equation}
The perturbations were done in the embedding space, instead of the text space, to be able to control its magnitude.
Each perturbed embedding is evolved deterministically under the same mapping $F$, yielding an ensemble of nearby trajectories in latent space.

By analyzing the divergence, recurrence structure, and effective dimensionality of these trajectories, we assess whether the internal dynamics of LLMs exhibit signatures of deterministic chaos. The quantitative diagnostics and experimental procedures used for this analysis are described in Sec.~\ref{sec:methods}.

\section{Methods}
\label{sec:methods}

This section describes the methodological framework used to characterize chaotic signatures in large language models (LLMs). Building on the dynamical-systems formalism introduced in Sec.~\ref{sec:preliminaries}, we analyze divergence, recurrence structure, and effective dimensionality of latent-state trajectories. Our analysis combines complementary trajectory distance metrics, Lyapunov exponent estimation, recurrence analysis, and fractal dimension estimation. We further complement these trajectory-level diagnostics with an exact, operator-level Jacobian analysis of the model's architectural sub-blocks (Sec.~\ref{subsec:computational_jacobian}). We conclude by detailing the experimental setup and data acquisition procedures.

\subsection{Trajectory Distance Metrics}
\label{subsec:trajectory_distance_metrics}

Let
\begin{equation}
\mathbf{X} = (\mathbf{x}_1,\dots,\mathbf{x}_n), \qquad
\mathbf{Y} = (\mathbf{y}_1,\dots,\mathbf{y}_m)
\end{equation}
denote two latent-state trajectories, where $\mathbf{x}_i,\mathbf{y}_j \in \mathbb{R}^D$. The length can differ ($n$ and $m$), since an LLM does not return a fixed-length answer; after reaching the token corresponding to the stopping, it ends answer generation. To quantify divergence between trajectories, we employ a suite of complementary distance metrics capturing both local (pointwise) and global (geometric) differences.

There is no canonical distance measure in high-dimensional latent or semantic spaces. Different metrics emphasize different geometric properties, such as orientation, alignment, or global shape. To ensure robustness and to avoid artifacts arising from a specific choice of metric, we therefore employ multiple complementary measures. Similar approaches have been used successfully in the analysis of vector representations of text and embeddings \cite{text2vec,reimers2019sentencebertsentenceembeddingsusing}.

\subsubsection{Vector-wise Metrics}
\label{sssec:methods_vector_metrics}

Vector-wise metrics compare corresponding points along two trajectories and yield a time series of distances. The primary metric we used is the \textbf{cosine distance},
\begin{equation}
d_{\mathrm{cos}}(\mathbf{x},\mathbf{y})
=
1 - \frac{\mathbf{x}\cdot\mathbf{y}}{\|\mathbf{x}\|\,\|\mathbf{y}\|},
\end{equation}
which measures angular dissimilarity between vectors. Cosine distance is widely used in embedding spaces and is well-suited for normalized hidden states. When vectors are normalized, cosine and Euclidean distances are monotonically related and yield qualitatively similar results.

\subsubsection{Trajectory-wise Metrics}
\label{sssec:methods_trajectory_metrics}

Trajectory-wise metrics treat trajectories as geometric objects rather than as pointwise-aligned sequences. These metrics yield a single scalar distance between two trajectory segments. To obtain time-resolved divergence curves, we apply them within a sliding window of increasing size (Sec.~\ref{sssec:methods_sliding_window}). Unless otherwise stated, Euclidean distance is used as the base metric between individual points.

\begin{itemize}
\item \textbf{Dynamic Time Warping (DTW)} \cite{SalvadorChan2007} computes an optimal nonlinear alignment between two sequences. Given a cost matrix $C_{i,j}=\|\mathbf{x}_i-\mathbf{y}_j\|$, the accumulated cost matrix satisfies
\begin{equation}
D_{i,j} = C_{i,j} + \min(D_{i-1,j},D_{i,j-1},D_{i-1,j-1}).
\end{equation}

\item \textbf{Discrete Fréchet Distance} \cite{Denaxas2023,EiterMannila1994} measures the minimal maximum separation between two curves under monotonic reparameterization,
\begin{equation}
d_F(X,Y) = \min_{\pi} \max_{(i,j)\in\pi} \|\mathbf{x}_i-\mathbf{y}_j\|,
\end{equation}
where $\pi$ ranges over all monotonic couplings of the sampled points.

\item \textbf{Hausdorff Distance} \cite{SciPyDirectedHausdorff} quantifies the maximum mismatch between two point sets,
{\small
\begin{equation}
d_H(X,Y) =
\max\!\left(
\sup_{\mathbf{x}\in X}\inf_{\mathbf{y}\in Y}\|\mathbf{x}-\mathbf{y}\|,
\sup_{\mathbf{y}\in Y}\inf_{\mathbf{x}\in X}\|\mathbf{x}-\mathbf{y}\|
\right).
\end{equation}
}

\item \textbf{Cross-Correlation Dissimilarity.}
This metric compares internal self-similarity structures. For each trajectory, a cosine self-similarity matrix $S_X$ is computed and flattened. The dissimilarity is defined as
\begin{equation}
d = 1 - \rho
=
1 - \frac{\operatorname{cov}(S_X^{\mathrm{flat}},S_Y^{\mathrm{flat}})}
{\sigma_{S_X^{\mathrm{flat}}}\sigma_{S_Y^{\mathrm{flat}}}},
\end{equation}
where $\rho$ is the Pearson correlation coefficient.
\end{itemize}

\subsubsection{Sliding Window Analysis}
\label{sssec:methods_sliding_window}

Trajectory-wise metrics yield a single scalar distance for a pair of trajectories. To obtain a timeseries of the divergence, we apply each metric within a sliding window that advances synchronously along both trajectories. Each window produces a scalar distance value, yielding a distance time series.

The use of sliding windows reflects the finite context window of autoregressive LLM inference, where the current state depends on a bounded history of previous states. For cosine distance, an optional windowed variant is used in which vectors within each window are averaged prior to pointwise comparison.

\subsection{Lyapunov Exponent Estimation}
\label{subsec:methods_lyapu}

To quantify sensitivity to initial conditions, we estimate maximal Lyapunov exponents from the divergence of nearby trajectories in latent space. A central challenge is that hidden-state trajectories are bounded due to normalization and extremely large number of dimensions, implying that inter-trajectory Euclidean distances inevitably saturate and generally extremely fast. Consequently, Lyapunov exponents can only be estimated from the initial, pre-saturation regime of divergence.

For each prompt, we generate an ensemble of perturbed trajectories as described in Sec.~\ref{subsec:init_conditions}. Distances are computed between corresponding hidden states of the final transformer layer or of the sentence embedding vectors using cosine distance. Because the effective dynamical state is distributed across the context window, we apply a sliding window of size $w=16$ (results were robust for $w\in\{8,32\}$). This yields a time series of distances $d(t)$.

For each trajectory pair, an exponential growth model,
\begin{equation}
d(t) \approx d_0 \, e^{\lambda t},
\end{equation}
is fitted over a contiguous interval starting at the first generated token and ending before saturation; in a chaotic system this yields a positive exponent. As shown in Sec.~\ref{res:lyapunov}, however, the observed divergence is dominated by discrete, jump-like events rather than by smooth exponential growth, so such fits generally attain only a low coefficient of determination. We therefore complement the finite-time exponent with a more robust characteristic quantity: the \textbf{first-divergence time}, the number of generated tokens until two trajectories first select a different token, whose distribution over an ensemble of perturbed trajectories serves as an inverse measure of chaoticity analogous to a Lyapunov time.

\subsection{Recurrence Analysis}
\label{subsec:recurrence_analysis_methods}

To characterize the geometric and temporal structure of trajectories, we employ recurrence analysis. Given a trajectory $\{\mathbf{x}_i\}_{i=1}^N$, we construct a recurrence matrix by thresholding the pairwise distance matrix,
\begin{equation}
R_{i,j}(\epsilon)
=
\Theta\!\left(\epsilon - d(\mathbf{x}_i,\mathbf{x}_j)\right),
\end{equation}
where $d(\cdot,\cdot)$ denotes the chosen distance metric and $\epsilon$ is a recurrence threshold.

Recurrence plots provide a qualitative visualization of dynamical structure. 
To ensure comparability across trajectories and representations, thresholds are selected such that the recurrence rate is held fixed, rather than fixing $\epsilon$ directly. This procedure controls for differences in scale and metric choice.

\subsection{Dimensionality Estimation}
\label{subsec:dimensionality_methods}

We estimate the correlation dimension $D_2$ of trajectories using the Grassberger--Procaccia algorithm \cite{GRASSBERGER1983189}. The correlation integral,
\begin{equation}
C(\epsilon)
=
\frac{1}{N(N-1)}
\sum_{i\neq j}
\Theta\!\left(\epsilon-\|\mathbf{x}_i-\mathbf{x}_j\|\right),
\end{equation}
measures the probability that two points on the trajectory are separated by less than $\epsilon$. For sufficiently small $\epsilon$, the correlation integral scales as
\begin{equation}
C(\epsilon) \sim \epsilon^{D_2}.
\end{equation}
The correlation dimension is obtained from the slope of the linear region in a log--log plot of $C(\epsilon)$ versus $\epsilon$. This method allows us to distinguish low-dimensional structured dynamics from high-dimensional stochastic behavior. Throughout, ``low-dimensional'' is to be understood relative to the ambient space: the diagnostic detects trajectories concentrating on a manifold whose effective dimension is orders of magnitude below the hidden dimension $D$.

\subsection{Computational Jacobian Analysis}
\label{subsec:computational_jacobian}

The diagnostics above characterize the global geometry of a trajectory but do not indicate \textbf{which} architectural components are responsible for the observed stretching and folding. To answer this, we complement the trajectory-level analysis with an exact, operator-level Jacobian analysis of the Transformer block, in the spirit of the mean-field tradition of studying stability through the layer Jacobian spectrum (Sec.~\ref{sec:prev_work}), but applied directly to a trained, autoregressive LLM rather than to a randomly initialized network. Because of the high dimensionality of the hidden state, direct computation and storage of a full-sequence Jacobian is memory-prohibitive; we instead extract intermediate activations via forward hooks on each sub-block during a standard forward pass and compute Jacobians locally, one sub-block at a time.

\subsubsection{Local Token-wise Operators}
\label{sssec:methods_jacobian_local}
For token-wise operators (the feed-forward block and the normalization layers), the Jacobian $\mathbf{J}(\mathbf{x}) = \partial \mathbf{f}(\mathbf{x})/\partial \mathbf{x} \in \mathbb{R}^{D \times D}$ acts independently on each token and can be computed exactly using vectorized reverse-mode automatic differentiation,
\begin{equation}
\mathbf{J}_{\text{chunk}}(\mathbf{x}) = \texttt{vmap}(\texttt{jacrev}(f))(\mathbf{x}),
\end{equation}
applied in chunks along the token dimension. From $\mathbf{J}(\mathbf{x})$ we extract the spectral norm $\|\mathbf{J}(\mathbf{x})\|_2 = \sigma_{\max}(\mathbf{J}(\mathbf{x}))$, which bounds the maximal local stretching of a perturbation along its most expansive direction, and the expected isotropic expansion
\begin{equation}
\bar\lambda = \frac{1}{D}\|\mathbf{J}(\mathbf{x})\|_F^2,
\end{equation}
which gives the average stretching factor applied to a randomly oriented perturbation; $\bar\lambda>1$ indicates net local phase-space expansion on average. Static weight metrics (e.g.\ scaled Frobenius norms of the projection matrices) are computed directly from the model parameters, while dynamic activation densities are obtained from the intermediate pre- and post-activation hidden states.

\subsubsection{Global Sequence Attention Jacobian}
\label{sssec:methods_jacobian_attn}
Self-attention couples all $N$ tokens in the context window and therefore cannot be reduced to a token-wise operator. We instead define the global sequence Jacobian $\mathbf{J}_{\text{Attn}} = \partial \mathbf{Y}/\partial \mathbf{X} \in \mathbb{R}^{(ND)\times(ND)}$, relating a perturbation of the full sequence matrix $\mathbf{X}\in\mathbb{R}^{N\times D}$ to the corresponding output perturbation. Materializing $\mathbf{J}_{\text{Attn}}$ explicitly requires $\mathcal{O}(N^2D^2)$ memory, which is not feasible even for moderate context lengths. We instead estimate its spectral norm $\|\mathbf{J}_{\text{Attn}}\|_2$ with a matrix-free power iteration that never constructs the full matrix: starting from a random unit-norm direction $\mathbf{v}_0\in\mathbb{R}^{N\times D}$, at each iteration $k$ we compute the pushforward (JVP) and pullback (VJP) of the attention operator directly on the automatic-differentiation graph,
\begin{align}
\mathbf{u}_k &= \mathbf{J}_{\text{Attn}}\mathbf{v}_k = \texttt{jvp}(\text{Attn},(\mathbf{X},),(\mathbf{v}_k,))_1,\\
\mathbf{w}_k &= \mathbf{J}_{\text{Attn}}^{T}\mathbf{u}_k = \texttt{vjp}(\text{Attn},\mathbf{X})(\mathbf{u}_k)_1,
\end{align}
and update $\mathbf{v}_{k+1}=\mathbf{w}_k/\|\mathbf{w}_k\|_F$. After $K=50$ iterations, the Rayleigh quotient
\begin{equation}
\sigma_{\max}(\mathbf{J}_{\text{Attn}}) \approx \sqrt{\frac{\langle \mathbf{w}_k,\mathbf{v}_k\rangle}{\langle \mathbf{v}_k,\mathbf{v}_k\rangle}}
\end{equation}
gives a numerically stable estimate of the dominant singular value of the global attention Jacobian, capturing perturbation growth mediated jointly across tokens and feature dimensions.

\subsection{Experimental Setup}
\label{subsec:experiments}

All experiments were conducted using the \texttt{DeepSeek-R1-Distill-Qwen-1.5B} model \cite{deepseek2025r1distillqwen1.5b}. The operator-level Jacobian and layer-wise perturbation analyses of Sec.~\ref{subsec:layer_perturbations} were additionally replicated on the architecturally distinct, 3.8-billion-parameter \texttt{Phi-4-mini-instruct} model~\cite{microsoft2025phi4minitechnicalreportcompact}, with qualitatively identical results~\cite{jaca2026thesis}. Inference was performed on a Tesla T4 GPU using greedy decoding (temperature $T=0$) to ensure deterministic generation. The context window length was set to 3096 tokens. Hidden states were extracted after each transformer block, yielding tensors of shape (layers, tokens, hidden dimension).

Perturbation magnitudes were chosen empirically to preserve topical coherence while inducing measurable divergence, with $r\in[2.5\times10^{-4},5\times10^{-4}]$ (for more details c.f. \ref{sec:experiments}).

Alongside hidden-state trajectories, we analyze trajectories constructed from sentence-level embeddings of the generated text. Generated outputs are segmented into sentences, each mapped to a fixed-dimensional vector using a pretrained sentence encoder, and treated as a trajectory in semantic space. Because sentence embeddings aggregate information over multiple tokens, they provide smoother, lower-noise representations than single-token hidden states.

Applying the same divergence and recurrence analyses reveals qualitatively similar signatures of sensitivity to initial conditions and recurrence structure in both spaces. This indicates that the observed behavior is not specific to a particular internal representation. Unless otherwise stated, results shown in the main text use the \texttt{all-mpnet-base-v2} encoder
\footnote{We compared the results using three different embedding models: \texttt{all-mpnet-base-v2}, \texttt{intfloat/e5-large-v2}, and \texttt{facebook/contriever}, with no significant differences between them.}.

\section{Results}
\label{sec:results}
This section presents empirical evidence that the internal dynamics of
LLMs exhibit key signatures associated with deterministic chaos. We
first examine, at the level of individual architectural sub-blocks,
how a local perturbation is amplified or suppressed as it propagates
across the layers of the model. We then analyze trajectory divergence
and Lyapunov exponents, study how the divergence of
trajectories depends on the size of the attention window,
characterize recurrence structure, and finally investigate the
effective dimensionality of hidden-state trajectories across layers.

\subsection{Propagation of Perturbations Across Layers}
\label{subsec:layer_perturbations}

Before examining how perturbations manifest in the generated trajectory, we first characterize, at the level of a single Transformer block, how a local perturbation is amplified or suppressed as it propagates forward. This decomposes the trajectory-level chaos reported in the remainder of this section into the contributions of four architectural components: self-attention, the feed-forward (MLP) block, normalization, and the residual connection.

\subsubsection{Feed-Forward (MLP) Block}
For an input hidden-state vector $\mathbf{x}\in\mathbb{R}^D$, the SwiGLU feed-forward block computes
\begin{equation}
\text{MLP}(\mathbf{x}) = W_{\text{down}}\big(\text{SiLU}(W_{\text{gate}}\mathbf{x})\odot(W_{\text{up}}\mathbf{x})\big),
\end{equation}
where $W_{\text{gate}},W_{\text{up}}\in\mathbb{R}^{D_{\text{ff}}\times D}$, $W_{\text{down}}\in\mathbb{R}^{D\times D_{\text{ff}}}$, $\text{SiLU}$ is the Sigmoid Linear Unit, and $\odot$ is the Hadamard product. Differentiating with respect to $\mathbf{x}$ gives the Jacobian
\begin{equation}
\mathbf{J}(\mathbf{x}) = W_{\text{down}}\big[\text{diag}(\mathbf{s}_x)W_{\text{up}} + \text{diag}(\mathbf{d}_x)W_{\text{gate}}\big],
\end{equation}
with $\mathbf{s}_x = \text{SiLU}(W_{\text{gate}}\mathbf{x})$ and $\mathbf{d}_x = (W_{\text{up}}\mathbf{x})\odot\text{SiLU}'(W_{\text{gate}}\mathbf{x})$ the gating and gating-sensitivity activations. We track the spectral norm $\|\mathbf{J}(\mathbf{x})\|_2$, the isotropic expansion $\bar\lambda = \|\mathbf{J}(\mathbf{x})\|_F^2/D$, the activation densities $\overline{s_x^2}=\|\mathbf{s}_x\|_2^2/D_{\text{ff}}$ and $\overline{d_x^2}=\|\mathbf{d}_x\|_2^2/D_{\text{ff}}$, and the scaled Frobenius norms $\tilde F_{\{\text{gate,up,down}\}}$ of the static weight matrices (Sec.~\ref{subsec:computational_jacobian}).

\begin{figure}[H]
\centering
\includegraphics[width=\linewidth]{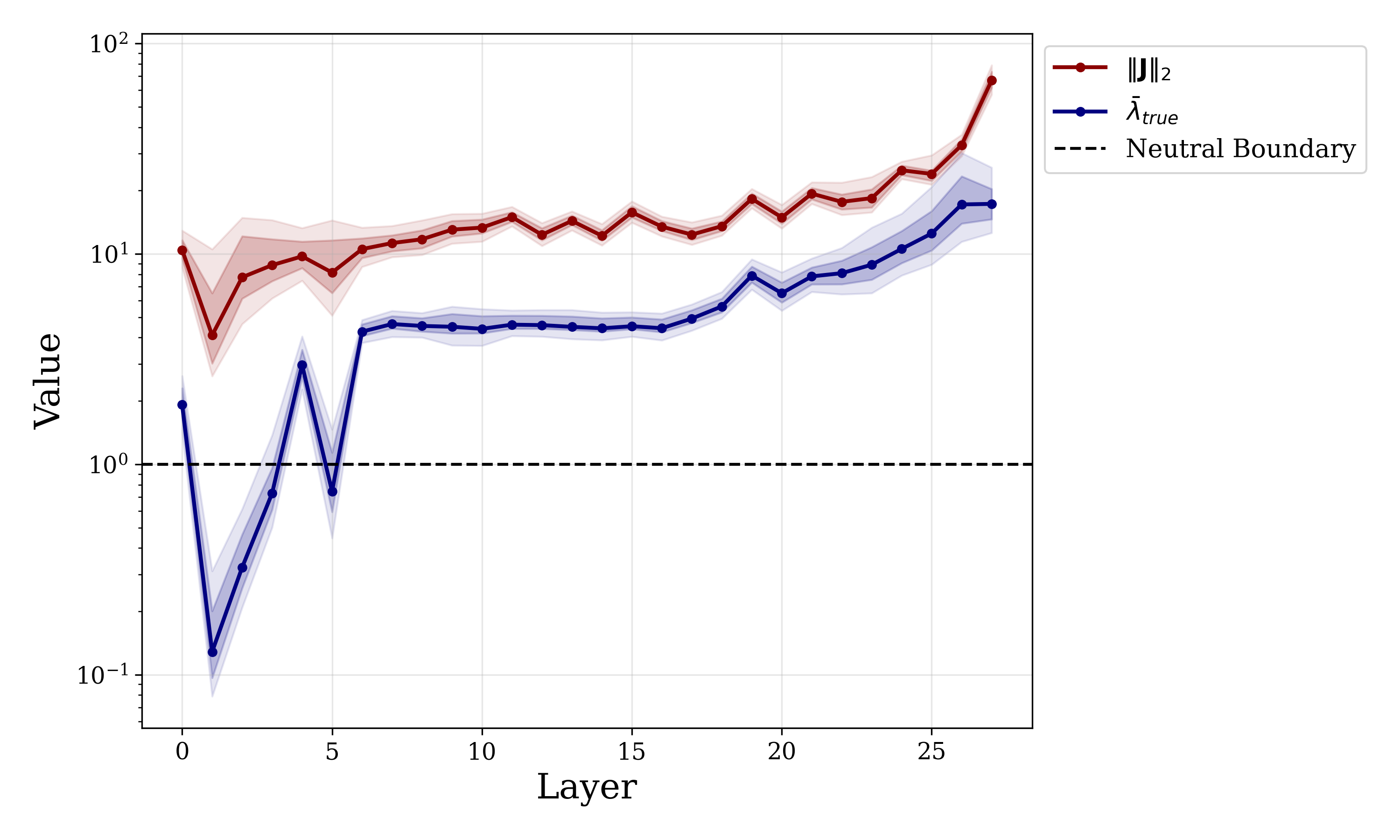}
\caption{Jacobian spectral norm $\|\mathbf{J}(\mathbf{x})\|_2$ and isotropic expansion $\bar\lambda$ of the feed-forward block, per layer. Shaded bands show the 10th--90th (light) and 25th--75th (dark) percentile ranges.}
\label{fig:jacobian_norm_and_lambda}
\end{figure}

The static weight Frobenius norms exceed $1$ at every layer, indicating the capacity to amplify perturbations. Consistent with this, activation densities and the measured Jacobian stretching (Fig.~\ref{fig:jacobian_norm_and_lambda}) both increase toward the final layers, where the dynamics are most active. The Jacobian spectral norm, $|\mathbf{J}(\mathbf{x})|_2$, exceeds $1$ throughout the network, and the average expansion factor, $\bar\lambda$, exceeds $1$ everywhere except the first few layers, confirming that the feed-forward block is generally expansive.

\paragraph{Mean-field prediction.} Expanding $\|\mathbf{J}(\mathbf{x})\|_F^2$ mixes the static weight matrices with the dynamic activations through cross-terms of the form $(W_{\text{down}}^TW_{\text{down}})\odot(W_{\text{up}}W_{\text{up}}^T)$. Neglecting off-diagonal correlations between the rows of the weight matrices and assuming their norms are approximately uniformly distributed, both standard mean-field simplifications in the study of random and trained networks alike~\cite{poole2016exponential_expressivity,schoenholz2017deep_information_propagation,yang2017meanfield_resnets_edge_of_chaos}, yields a closed-form estimate for the isotropic expansion that requires only the static weight norms and the activation densities (full derivation in Appendix~\ref{app:mft}):
\begin{equation}
\bar\lambda \approx \tilde F_{\text{down}}\left[\tilde F_{\text{gate}}\,\overline{d_x^2} + \tilde F_{\text{up}}\,\overline{s_x^2}\right].
\end{equation}

\begin{figure}[H]
\centering
\includegraphics[width=\linewidth]{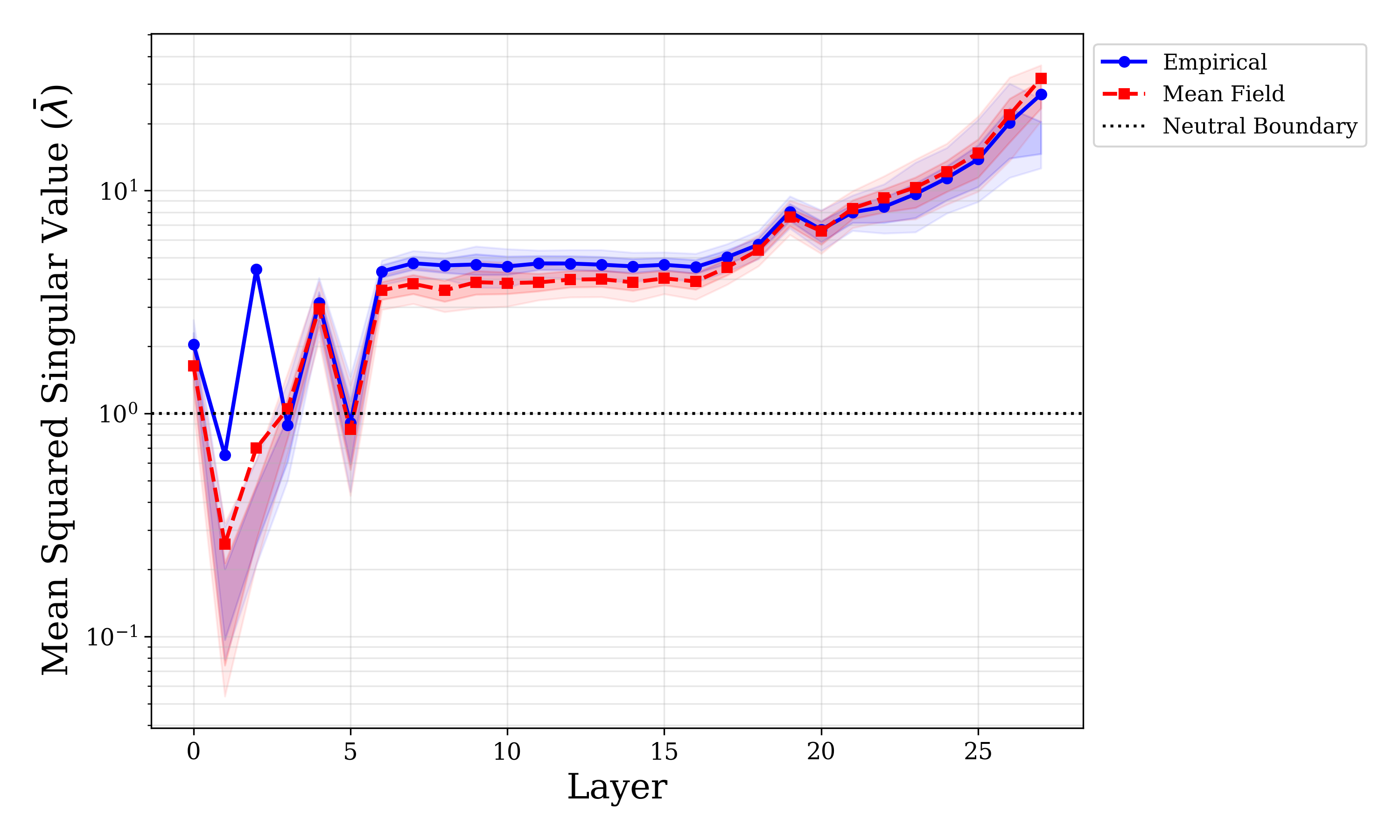}
\caption{Measured $\bar\lambda$ compared with the value predicted by the mean-field formula above. Shaded bands as in Fig.~\ref{fig:jacobian_norm_and_lambda}.}
\label{fig:mft}
\end{figure}

Fig.~\ref{fig:mft} shows that this simple closed-form estimate tracks the exact, empirically measured expansion factor across all layers, with the main discrepancy located in the second and third layers, where outlier tokens whose perturbation aligns with the dominant Jacobian direction inflate the (outlier-sensitive) mean used for this comparison. The assumptions underlying the formula are validated numerically in Appendix~\ref{app:mft}.

\subsubsection{Attention}
Let $\mathbf{X}\in\mathbb{R}^{N\times D}$ be the input sequence of $N$ tokens. A single attention head computes $\mathbf{Y}=\mathbf{A}\mathbf{X}W_VW_O$, with $\mathbf{A}=\text{softmax}(\mathbf{Z})$ applied row-wise to the causally masked logits $\mathbf{Z} = \mathbf{X}W_QW_K^T\mathbf{X}^T/\sqrt{d_k} + \mathcal{M}$. Differentiating $\mathbf{Y}$ with respect to a perturbation $\delta\mathbf{X}$ splits the response into two qualitatively different terms,
\begin{equation}
\delta\mathbf{Y} = \underbrace{(\delta\mathbf{A})\mathbf{X}W_VW_O}_{\text{nonlinear routing}} + \underbrace{\mathbf{A}(\delta\mathbf{X})W_VW_O}_{\text{static linear mixing}}.
\end{equation}
Because $\mathbf{A}$ is row-stochastic, the second, static term is a convex combination of the perturbed value vectors and, by the triangle inequality, cannot itself amplify the perturbation norm beyond the projection weight norm $\|W_VW_O\|_2$; the attention matrix can only ever \textbf{mix}, not \textbf{stretch}, through this term. All state-dependent amplification instead originates in the first, nonlinear term, whose logit perturbation $\delta\mathbf{Z}$ is bilinear in $\delta\mathbf{X}$ and $\mathbf{X}$, making $\delta\mathbf{Y}_{\text{route}}$ quadratic in $\mathbf{X}$ and hence genuinely nonlinear. The relevant softmax Jacobian $\mathbf{J}_{\text{sm}}(\mathbf{a}_i)=\text{diag}(\mathbf{a}_i)-\mathbf{a}_i\mathbf{a}_i^T$ vanishes only if a row of $\mathbf{A}$ saturates to a one-hot vector; otherwise it obeys the state-independent bound $\|\mathbf{J}_{\text{sm}}\|_2\le 1/2$~\cite{nair2025softmax12lipschitztightbound,emadi2026exactattentionsensitivitygeometry}, consistent with the more general non-Lipschitz behavior of self-attention on unbounded domains~\cite{kim2021lipschitzconstantselfattention}. We measure the attention-entropy ratio (entropy normalized by its causal maximum) across all heads and layers and find it fluctuating between $0.4$ and $0.8$: attention does not one-hot saturate, so the nonlinear routing term remains active throughout inference, and entropy collapse~\cite{zhai2023attention_entropy_collapse} is not observed at inference time in this model.

\begin{figure}[H]
\centering
\includegraphics[width=\linewidth]{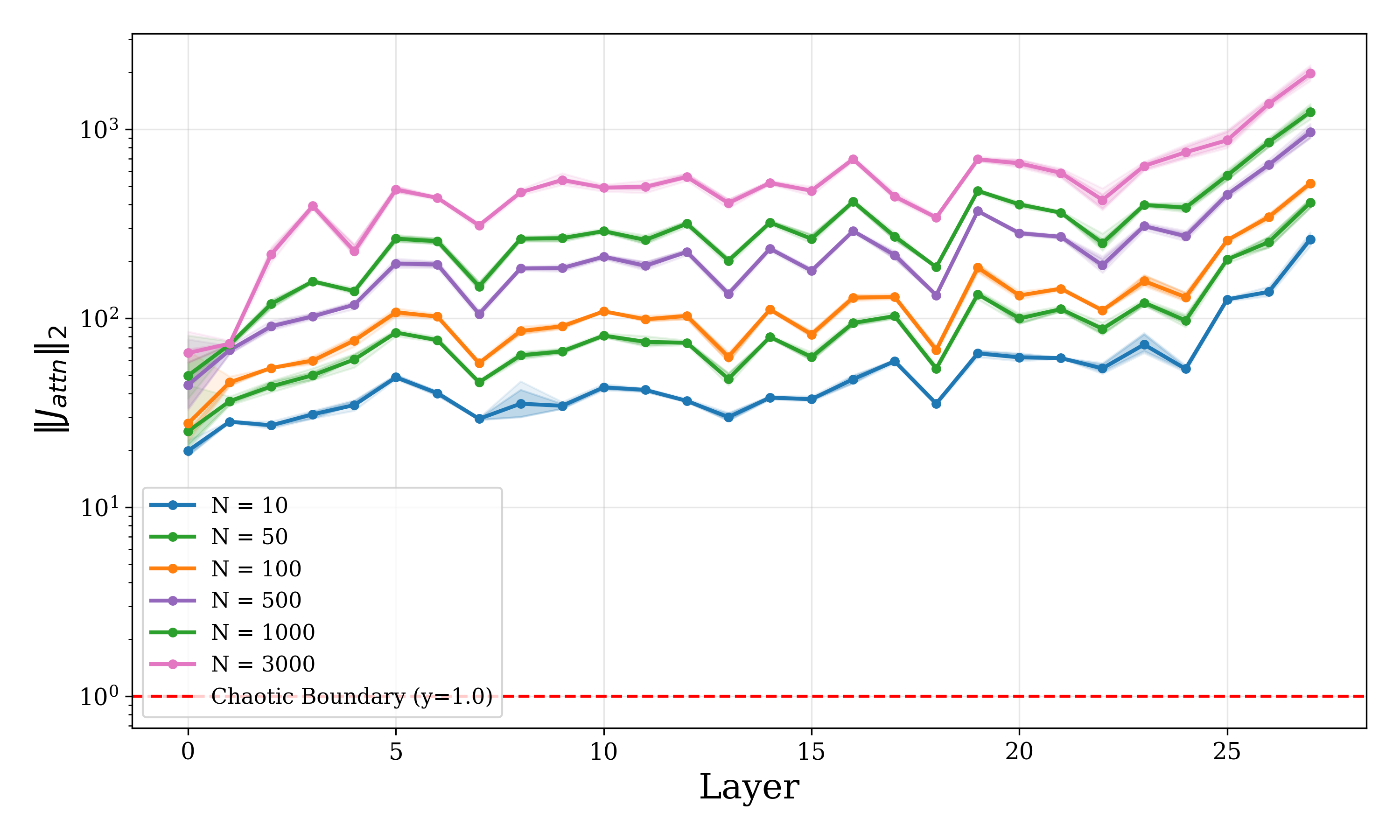}
\caption{Global sequence attention Jacobian spectral norm $\|\mathbf{J}_{\text{Attn}}\|_2$ (Sec.~\ref{sssec:methods_jacobian_attn}), per layer.}
\label{fig:attn_jacobian_spectral_norm}
\end{figure}

Because the causal mask lets a perturbation injected at an early token re-enter, through the key/value cache, the output of every later token, we also evaluate the full \textbf{global} sequence Jacobian $\mathbf{J}_{\text{Attn}}$ via the power iteration of Sec.~\ref{sssec:methods_jacobian_attn}. Its spectral norm is consistently positive and larger than $1$ at every layer (Fig.~\ref{fig:attn_jacobian_spectral_norm}), and, consistent with a bilinear coupling acting on an effectively isotropic, high-dimensional sequence, its magnitude scales approximately with $\sqrt{N}$. The static routing and mixing weight norms $\|W_QW_K^T\|_2$ and $\|W_VW_O\|_2$ are likewise found to be larger than $1$ at every layer, confirming that self-attention, like the feed-forward block, possesses the raw capacity for perturbation expansion.

\subsubsection{Normalization}
RMSNorm computes $\mathbf{y}=\mathbf{x}/\text{RMS}(\mathbf{x})$ with $\text{RMS}(\mathbf{x})=\|\mathbf{x}\|_2/\sqrt{D}$, whose Jacobian is
\begin{equation}
\mathbf{J}_{\text{RMS}} = \frac{\sqrt{D}}{\|\mathbf{x}\|_2}\left(\mathbf{I}-\frac{\mathbf{x}\mathbf{x}^T}{\|\mathbf{x}\|_2^2}\right) = \frac{1}{S}\mathbf{P}_\perp,
\end{equation}
an orthogonal projection $\mathbf{P}_\perp$ (which annihilates the radial component of any perturbation, $\mathbf{P}_\perp\mathbf{x}=\mathbf 0$) scaled by the inverse magnitude $S=\|\mathbf{x}\|_2/\sqrt D$ of the state. LayerNorm has the same structure acting on the mean-centered state. Both mechanisms actively counteract the expansion identified above: the projection removes exactly the radially outward component that unbounded stretching would produce, while the scaling means that if a preceding operator has already inflated $\|\mathbf{x}\|_2$, the very next normalization step shrinks the perturbation fed into the following nonlinear block by that same factor. Normalization thus supplies the ``folding'' counterpart to the stretching identified in the attention and feed-forward blocks\footnote{This also explains why cosine-based metrics, insensitive to the (normalized-away) radial direction, are particularly well suited to LLM hidden states, cf.\ Sec.~\ref{subsec:trajectory_distance_metrics}.}.

\subsubsection{Residual Connections}
Every sub-block $\mathbf{F}(\mathbf{x})=\text{MLP/Attn}(\text{Norm}(\mathbf{x}))$ is wrapped in a residual connection, $\mathbf{x}_{\text{res}}=\mathbf{x}+\mathbf{F}(\mathbf{x})$, so that the block Jacobian is $\mathbf{J}_{\text{res}}=\mathbf{I}+\mathbf{J}_F$. Unlike the eigenvalues, which merely shift by $1$, the singular values relevant to a single application of the operator do not follow a simple closed form, since $\|\mathbf{J}_{\text{res}}\delta\mathbf{x}\|_2^2 = \|\delta\mathbf{x}\|_2^2 + \|\mathbf{J}_F\delta\mathbf{x}\|_2^2 + 2\,\delta\mathbf{x}^T\mathbf{J}_F\delta\mathbf{x}$. Functionally, however, the identity branch is what allows perturbations to persist: it preserves precisely the radial component that the normalization projection $\mathbf{P}_\perp$ removed from the input to $\mathbf{F}$, carrying it forward along the residual stream even as normalization keeps the input to each nonlinear block bounded. This interplay, expansion in Attention/MLP, contraction in normalization, and information-preserving propagation through the residual stream, is what allows a global perturbation to persist and eventually grow while individual normalized sub-blocks never see an unbounded input, mirroring the way skip connections were shown to soften the order-to-chaos transition of purely feedforward networks~\cite{yang2017meanfield_resnets_edge_of_chaos}.

\subsubsection{Finite-Size Perturbations}
The derivations above use exact, infinitesimal derivatives evaluated one operator at a time. To see how a \textbf{finite} perturbation propagates through a full layer (attention, MLP, normalization, and residual connections acting together), we generate an ensemble of initial conditions on a small simplex around a single token (Sec.~\ref{subsec:init_conditions}) and track the perturbation magnitude across layers.

\begin{figure}[H]
\centering
\includegraphics[width=\linewidth]{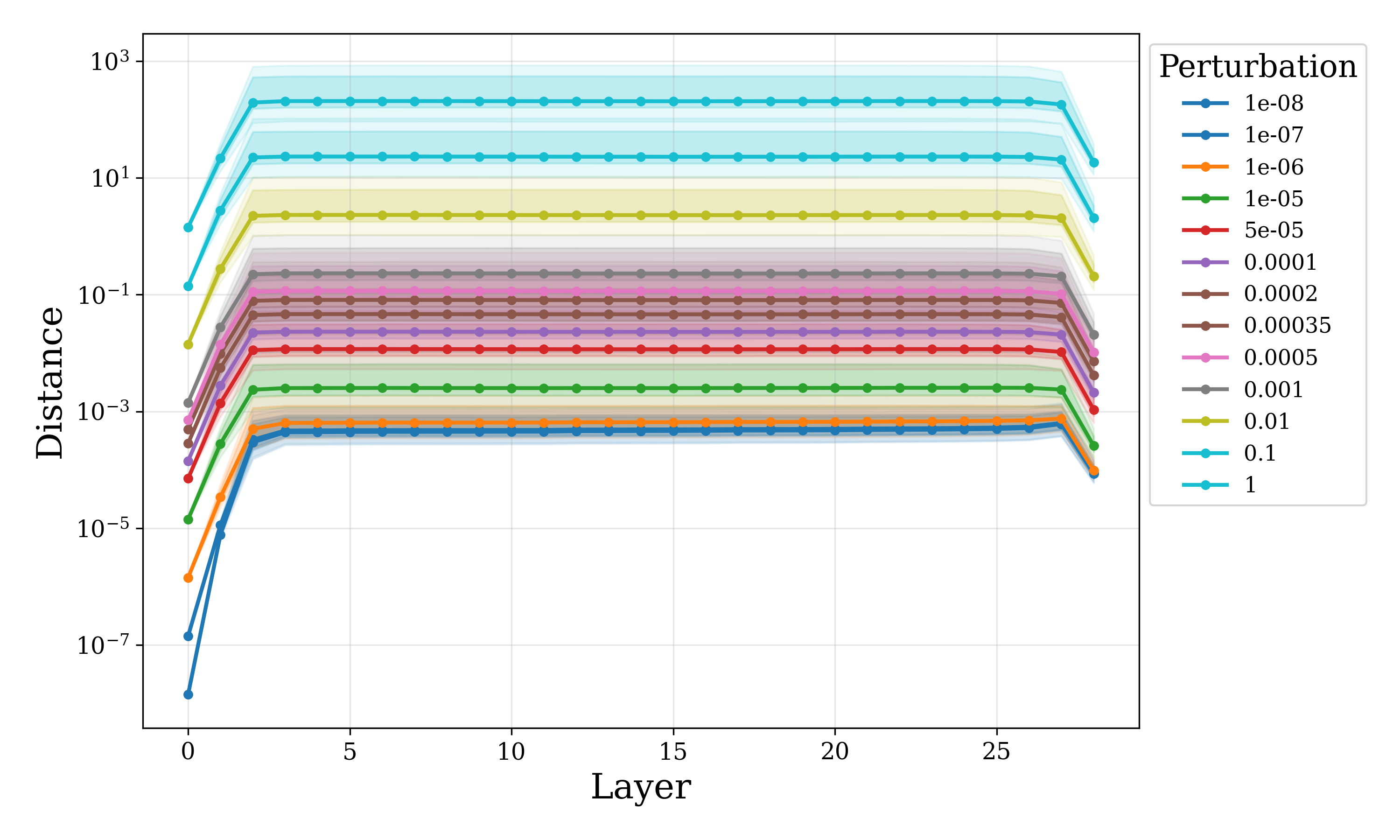}
\caption{Propagation of a finite perturbation across the layers of the model. Shaded bands as in Fig.~\ref{fig:jacobian_norm_and_lambda}.}
\label{fig:perturbation_layers}
\end{figure}

\begin{figure}[H]
\centering
\includegraphics[width=\linewidth]{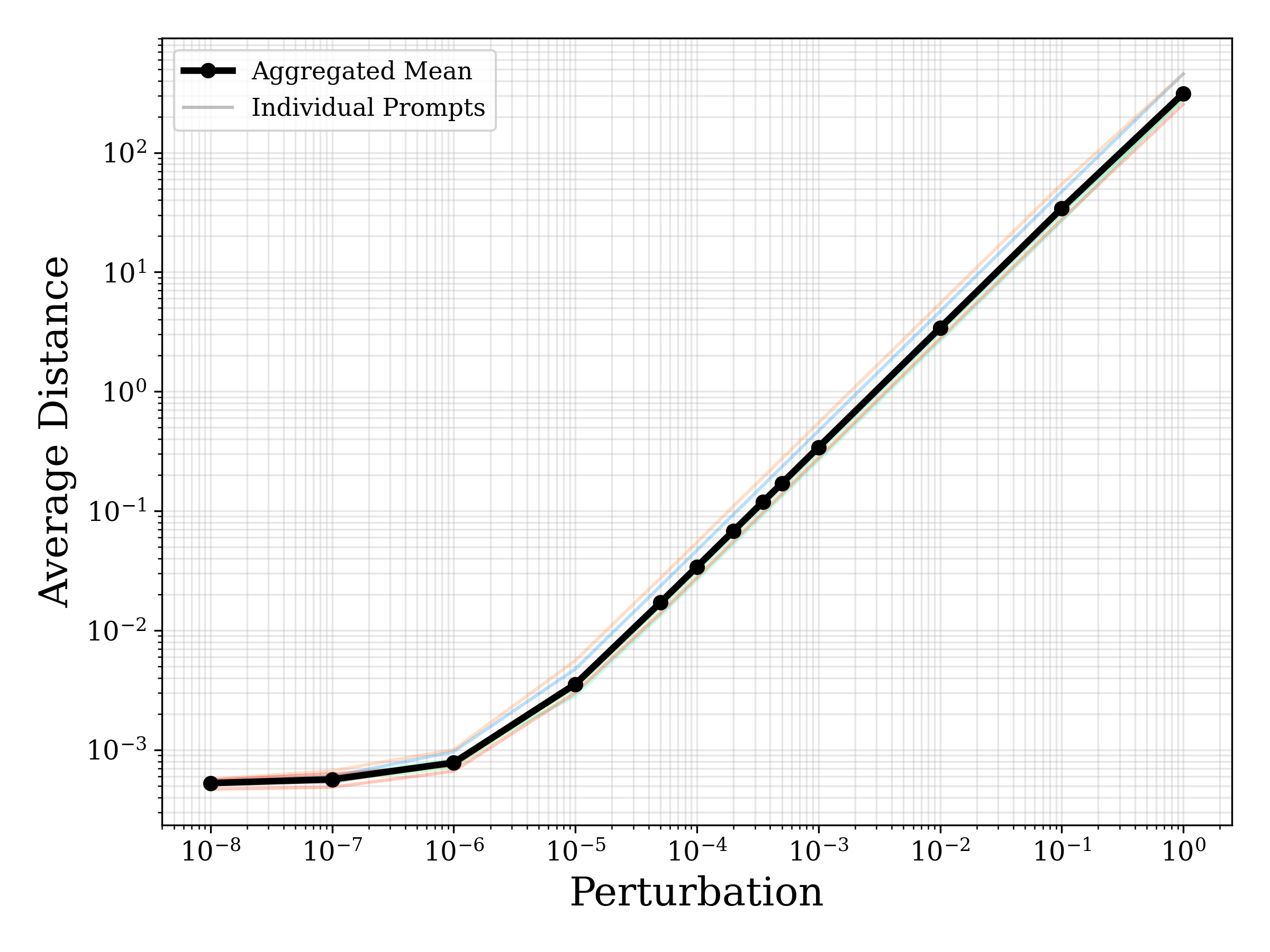}
\caption{Perturbation magnitude at each layer as a function of the initial perturbation radius $r$, on a log--log scale; the fitted exponent, $0.993\pm0.007$, is consistent with the linear-response exponent of $1$.}
\label{fig:scaling}
\end{figure}

The perturbation magnitude grows rapidly in the first few layers, plateaus through the middle of the network, and decreases again in the last layers (Fig.~\ref{fig:perturbation_layers}). We interpret the plateau as the regime in which the model expands the phase space to maximally mix contextual information, while the late-layer contraction reflects the network committing to a sharper next-token distribution ahead of the final unembedding. In the plateau region, the perturbation magnitude follows a power law in the initial perturbation radius $r$, with a fitted exponent of $0.993\pm 0.007$, matching the exponent of $1$ expected from the linearized tangent map $\mathbf{T}_{0\to\ell}=\prod_{k=1}^{\ell}\mathbf{J}_k$: $\|\delta\mathbf{x}_\ell\|_2\approx r\|\mathbf{T}_{0\to\ell}\hat{\mathbf u}\|_2$. The constant term of this fit recovers the aggregate per-layer Jacobian product, consistent in magnitude with the operator-level spectral norms of Figs.~\ref{fig:jacobian_norm_and_lambda} and~\ref{fig:attn_jacobian_spectral_norm}. At sufficiently small $r$, the power law breaks down into a constant perturbation floor set by the numerical precision of the underlying FP16 computation (activation values are typically concentrated in $[0.5,1.0]$, whose FP16 spacing is $2^{-11}\approx4.88\times10^{-4}$, matching the observed floor). The resulting picture, expansion in the middle of the network followed by late-layer contraction and a commitment to the output distribution, parallels the non-normal-to-symmetric spectral transition and cumulative low-rank bottleneck recently reported across the depth of full production LLMs~\cite{fernando2026transformer_residual_stream_dynamics}.

\subsection{Trajectory Divergence and Lyapunov Exponents}
\label{res:lyapunov}

We next examine the qualitative structure of the distance–time curves obtained from sentence-embedding trajectories. From a finite-time perspective, the slope of the logarithmic distance growth in the pre-saturation regime provides an estimate of the maximal Lyapunov exponent,
\begin{equation}
    \lambda_\mathrm{max}\simeq\frac{1}{\Delta t}\frac{d}{dt}\log[d(t)]
\end{equation}
where $d(t)$ is the trajectory separation measured in embedding space.

\begin{figure}[H]
    \centering
    \begin{subfigure}[t]{0.225\textwidth}
        \includegraphics[width=\linewidth]{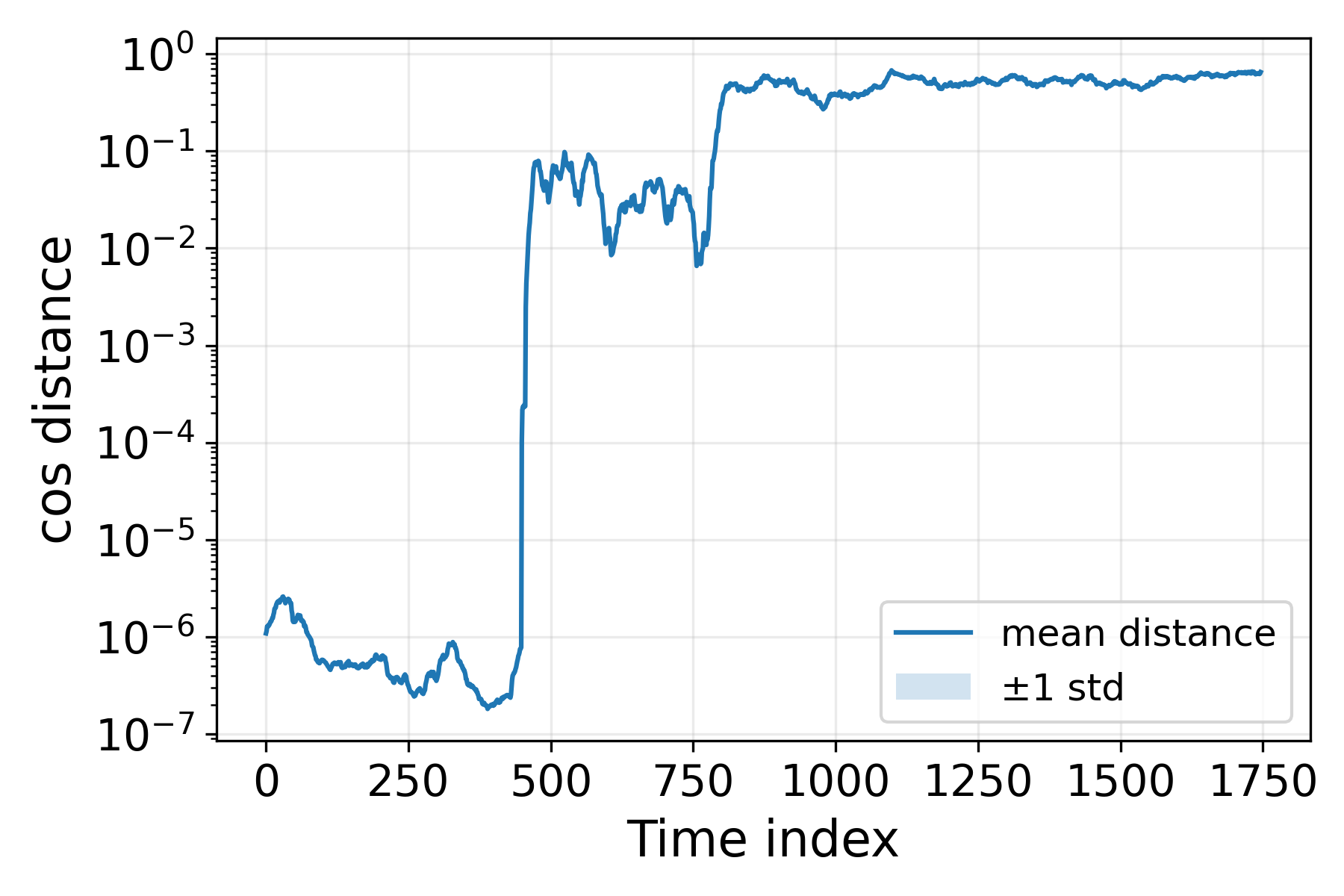}
        \caption{Single pair}
    \end{subfigure}\hfill
    \begin{subfigure}[t]{0.225\textwidth}
        \includegraphics[width=\linewidth]{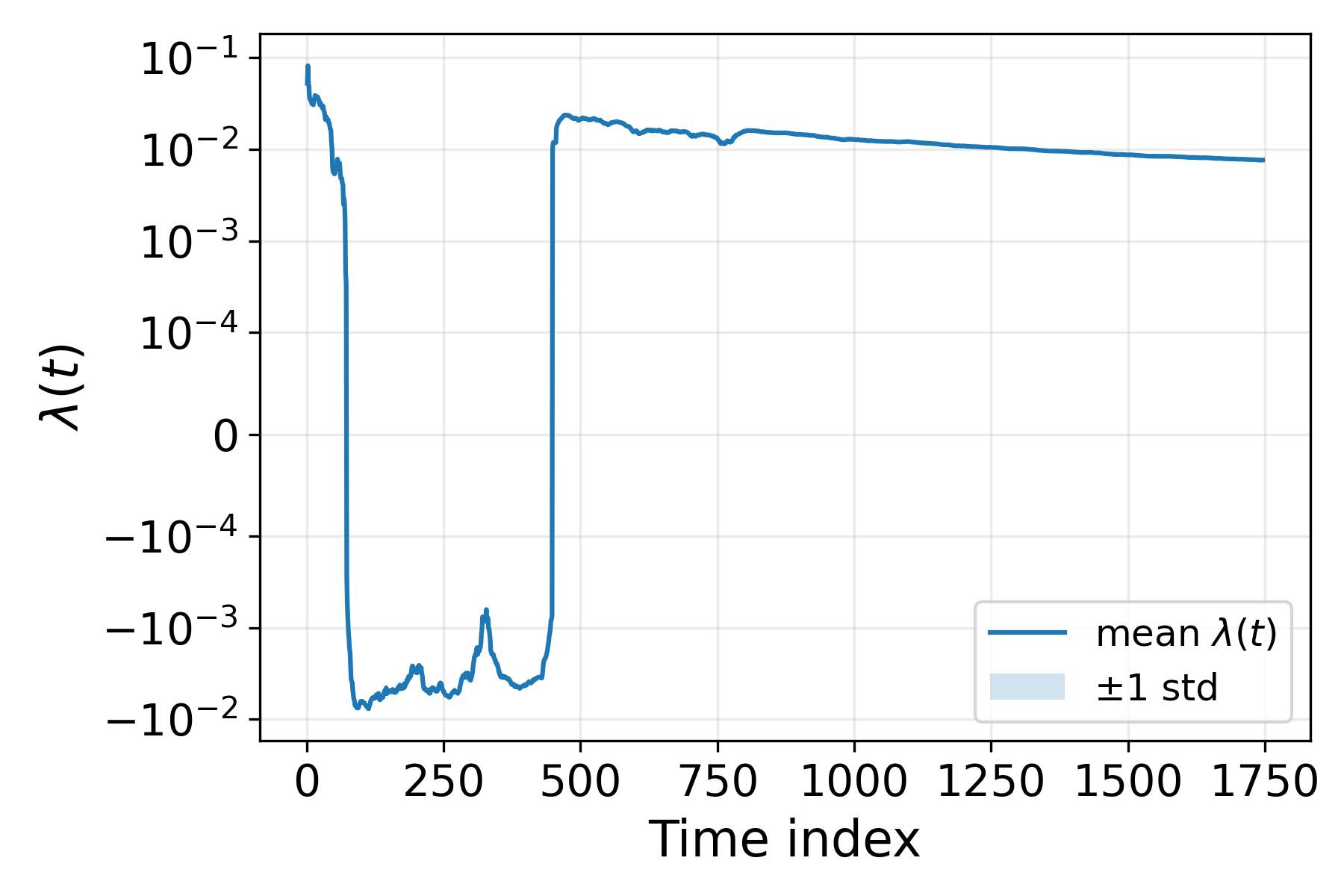}
    \end{subfigure}
    \\[0.5em]
    \begin{subfigure}[t]{0.225\textwidth}
        \includegraphics[width=\linewidth]{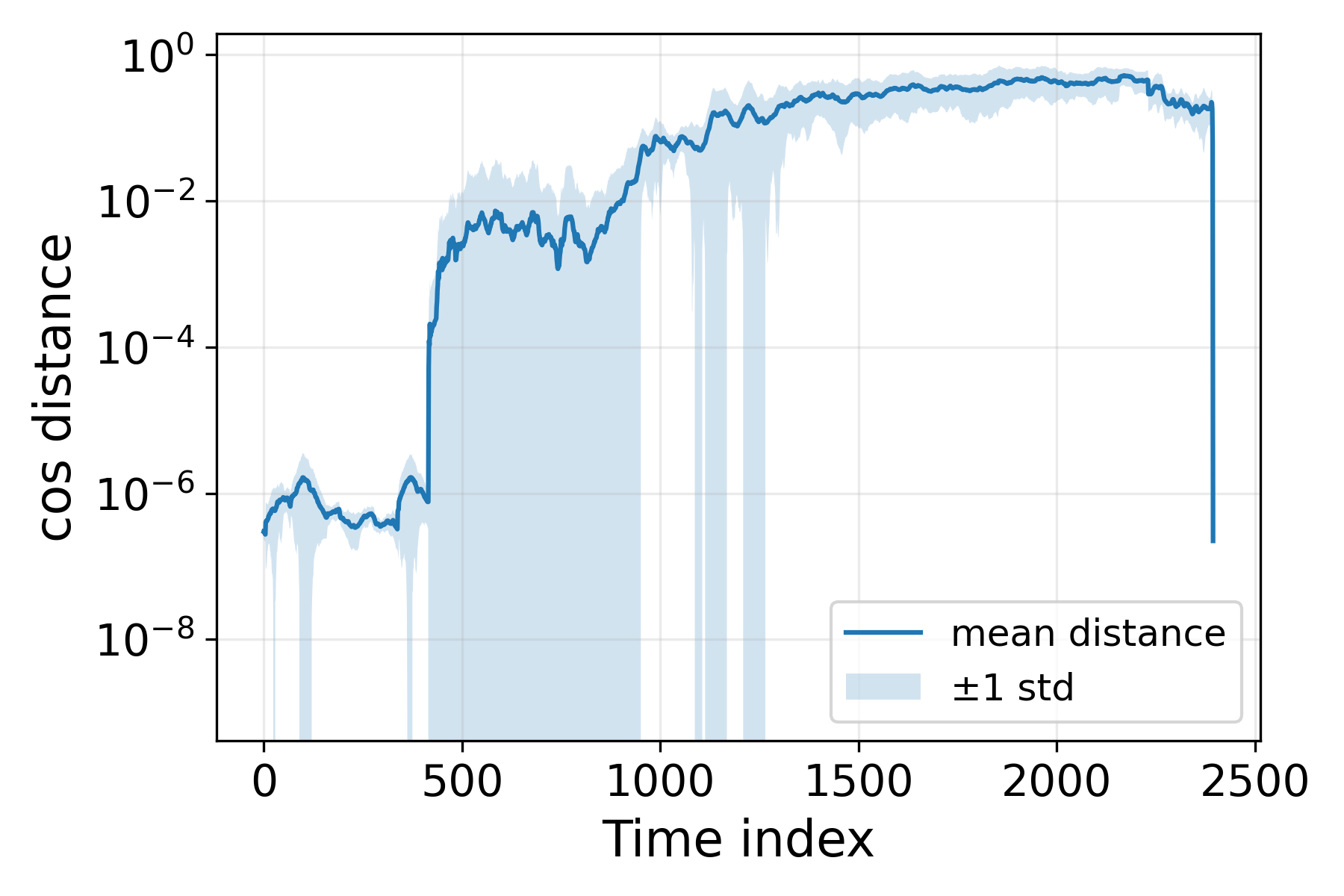}
    \end{subfigure}\hfill
    \begin{subfigure}[t]{0.225\textwidth}
        \includegraphics[width=\linewidth]{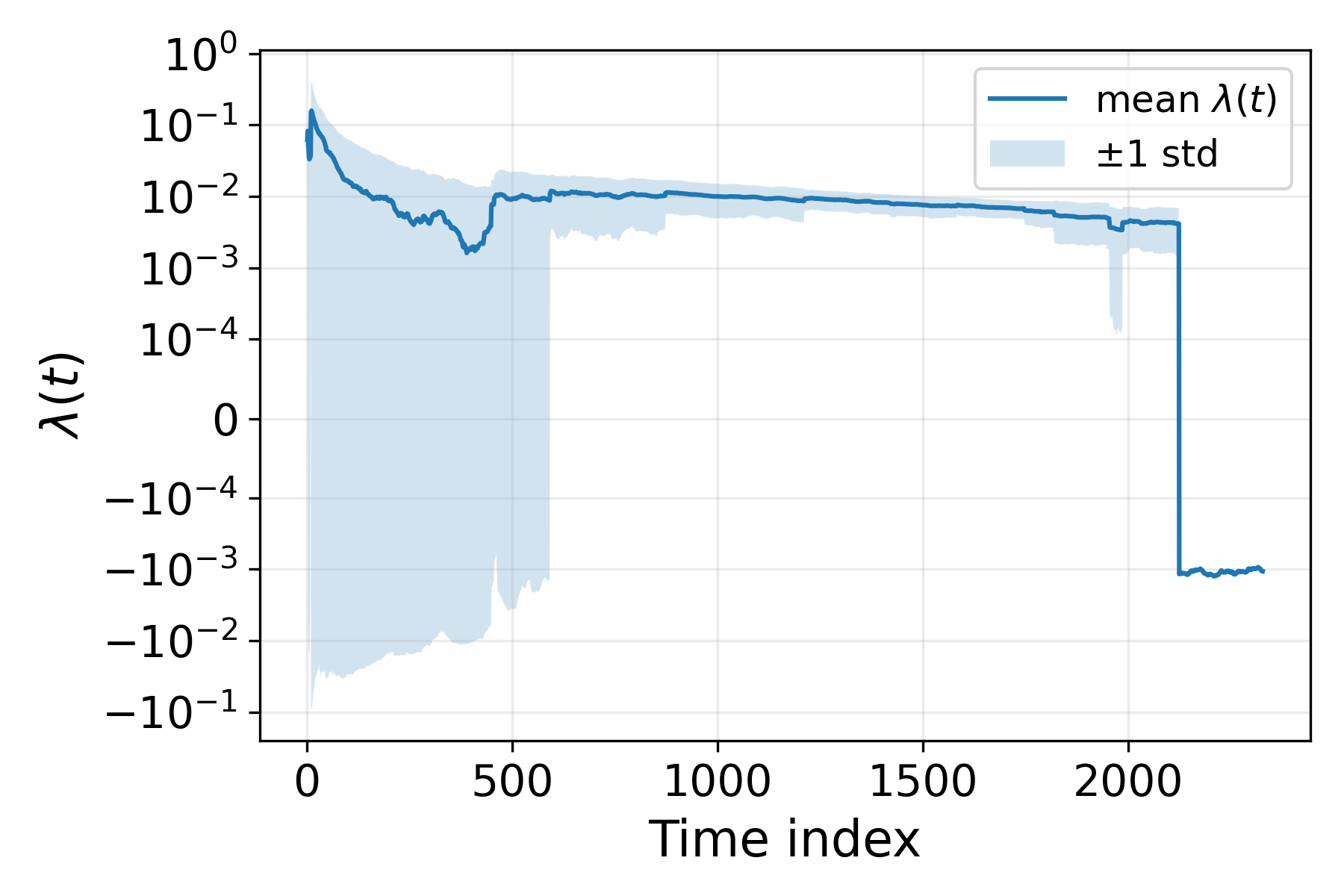}
        \caption{Ensemble mean}
    \end{subfigure}
    \caption{Cosine Distance of the hidden states and Lyapunov timeseries.  The maximal Lyapunov exponent is negative for the initial part, which corresponds to the tokens being the same, and the hidden state distance decreasing, although the perturbation remains latent in the hidden states. Eventually the selected token is different between the two trajectories, making the distance quickly increase and the lyapunov exponent to become positive. This triggers the trajectories to quickly diverge.}
    \label{fig:cos_dist_lyapunov_timeseries}
\end{figure}

\begin{figure}[H]
    \centering
    \begin{subfigure}[t]{0.225\textwidth}
        \includegraphics[width=\linewidth]{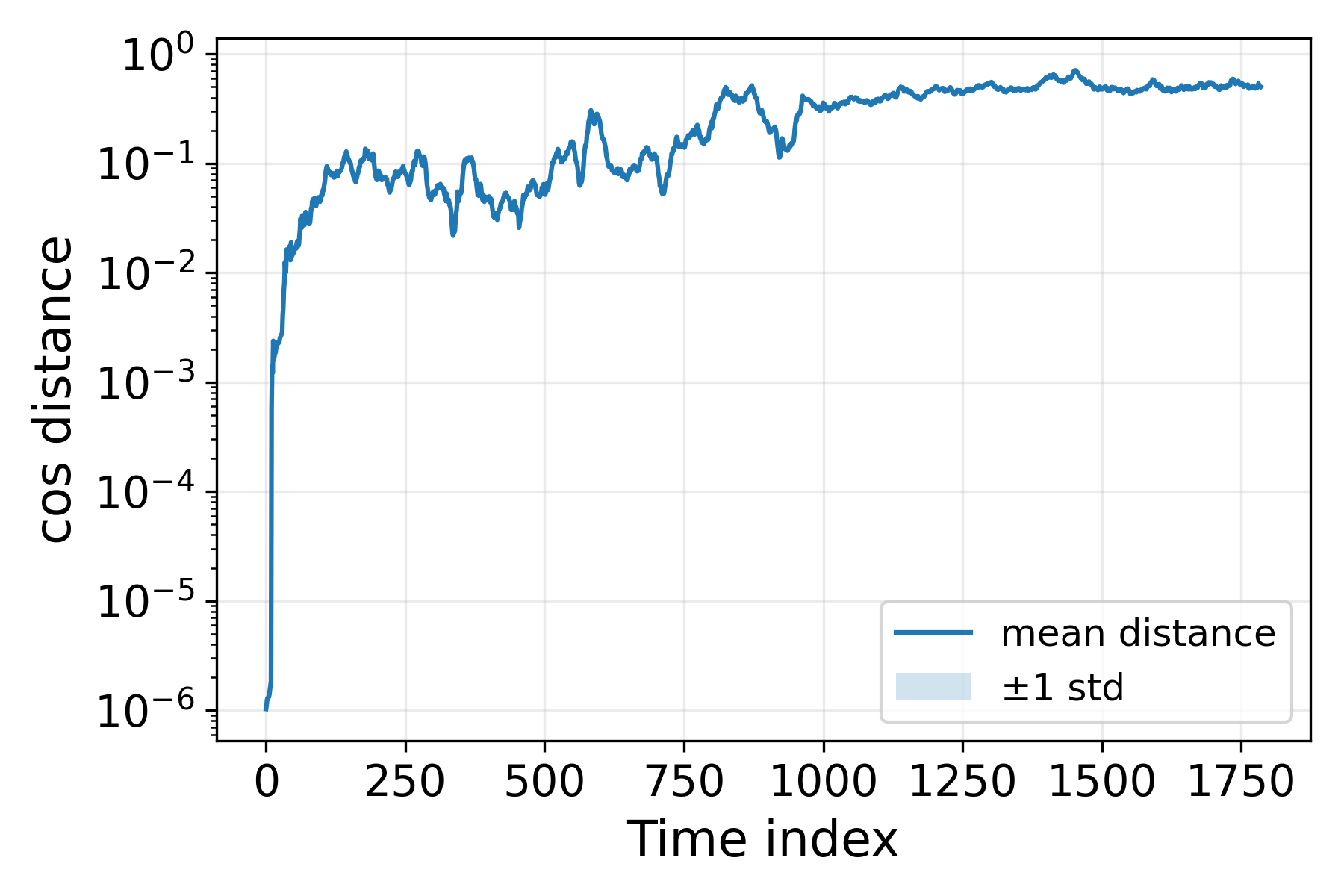}
        \caption{Outlier pair 1}
    \end{subfigure}\hfill
    \begin{subfigure}[t]{0.225\textwidth}
        \includegraphics[width=\linewidth]{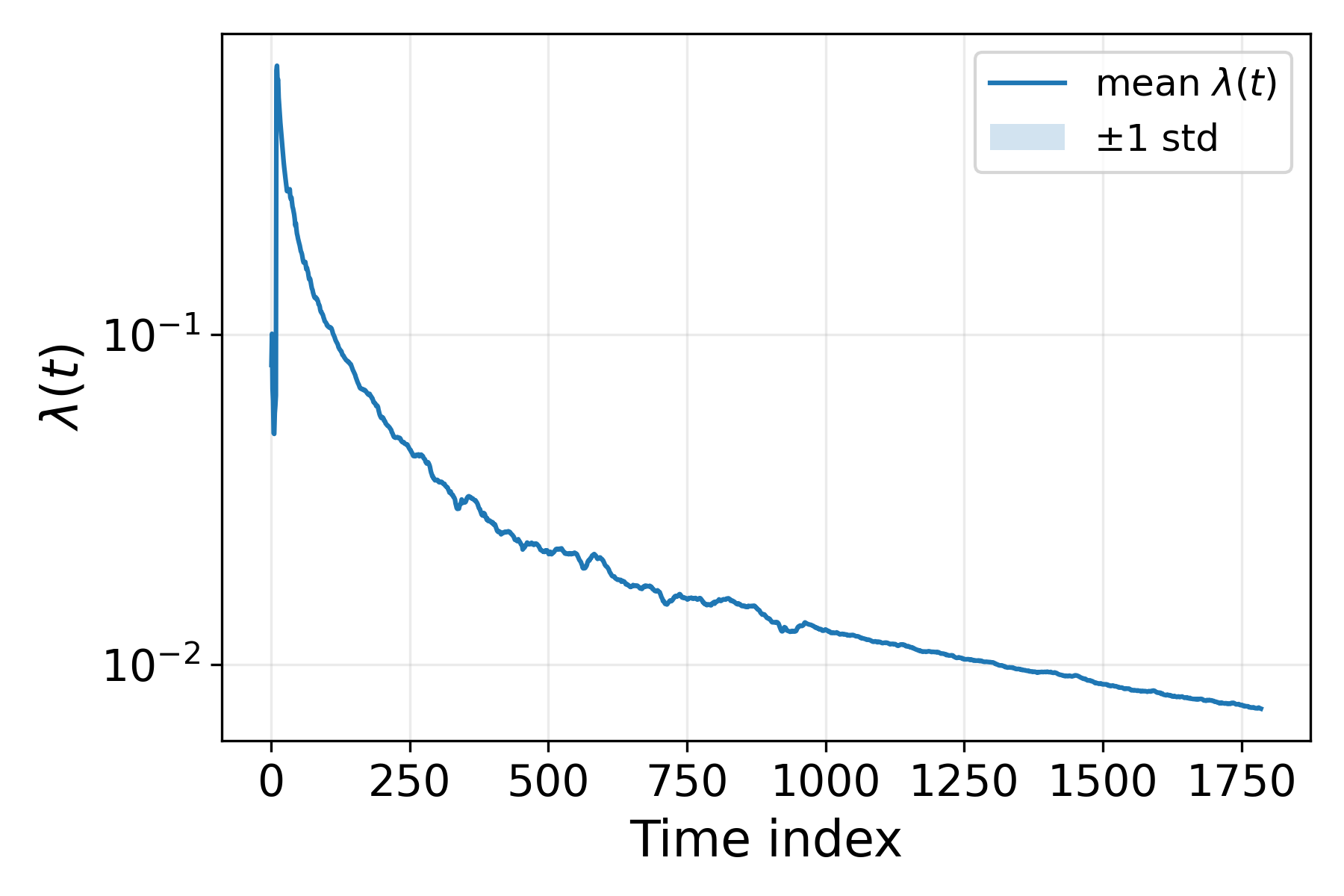}
    \end{subfigure}
    \\[0.5em]
    \begin{subfigure}[t]{0.225\textwidth}
        \includegraphics[width=\linewidth]{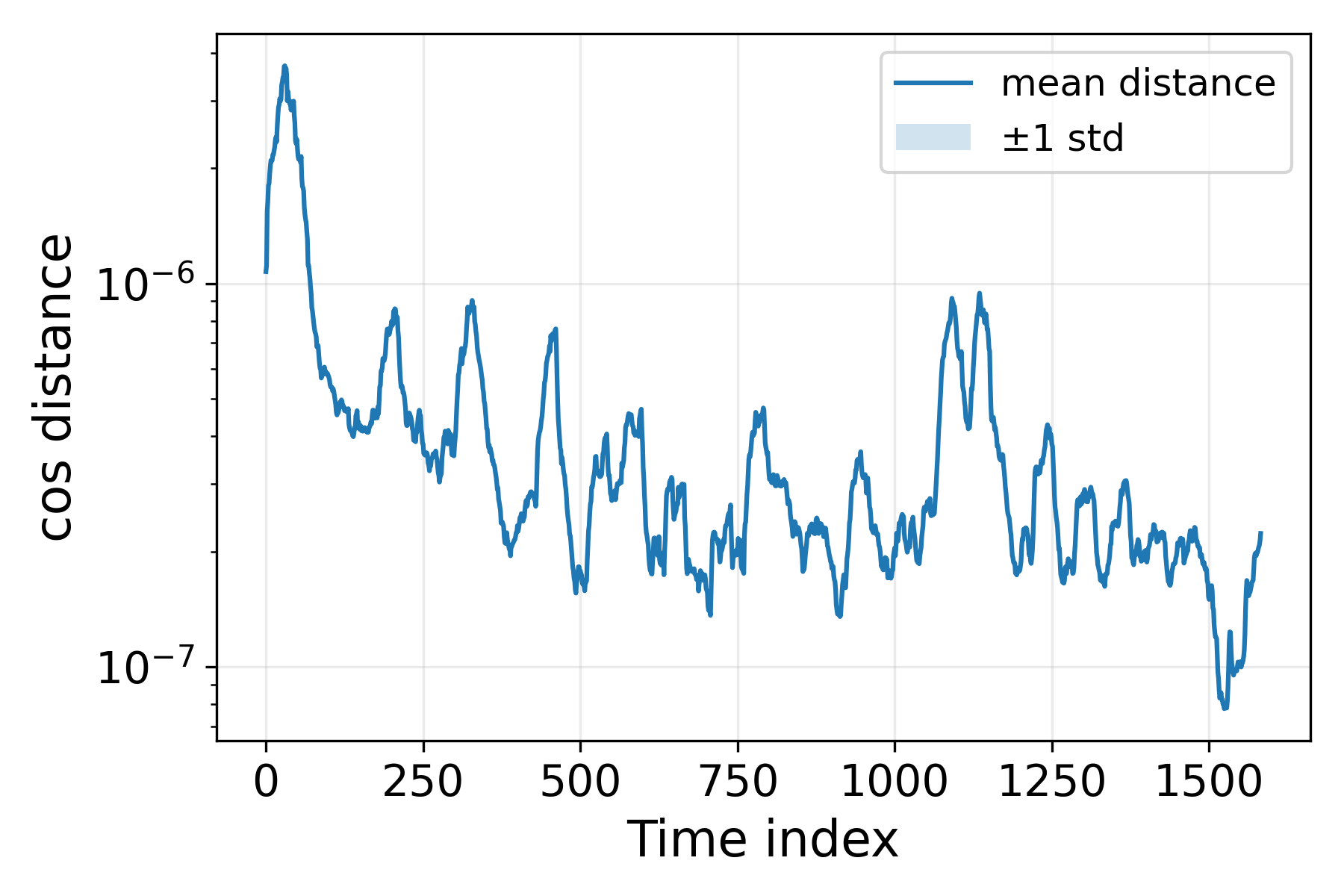}
        \caption{Outlier pair 2}
    \end{subfigure}\hfill
    \begin{subfigure}[t]{0.225\textwidth}
        \includegraphics[width=\linewidth]{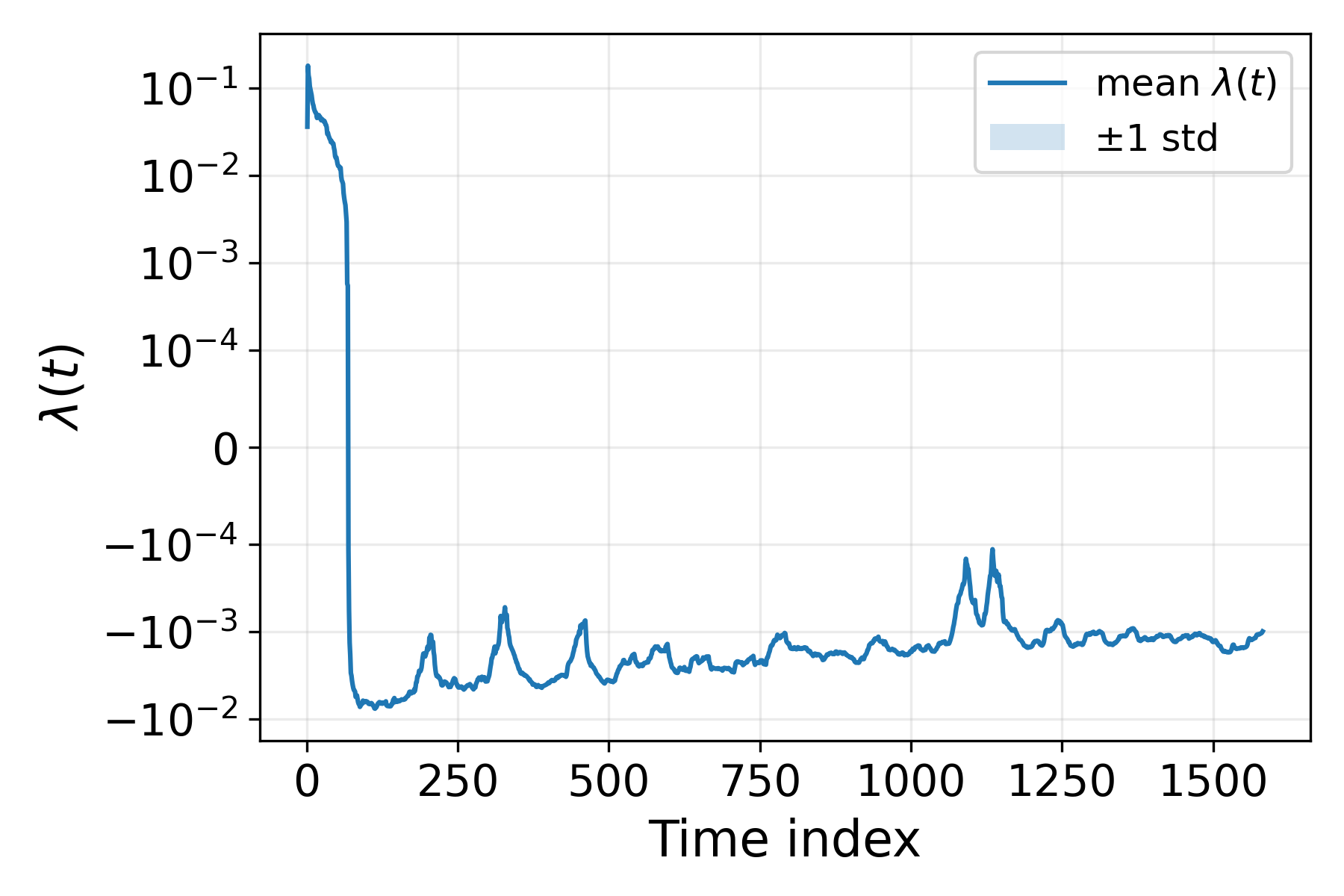}
    \end{subfigure}
    \caption{Examples of outlier trajectory pairs (roughly $10\%$ of the ensemble) whose divergence profiles deviate from the dominant latent-then-jump pattern of Fig.~\ref{fig:cos_dist_lyapunov_timeseries}.}
    \label{fig:cos_dist_lyapunov_outliers}
\end{figure}

The typical divergence profile (Fig.~\ref{fig:cos_dist_lyapunov_timeseries}) directly reflects the discrete token bottleneck of autoregressive generation. Immediately after the perturbation is applied, both trajectories select identical tokens: the perturbation remains \textbf{latent}, detectable only in the hidden states, where the measured distance may even shrink and the finite-time exponent is negative. At some step the accumulated perturbation, propagated and amplified through the nonlinear attention coupling, crosses the sampling threshold and a different token is selected; this single discrete difference feeds back into the context of all subsequent predictions and the trajectories separate rapidly until the distance saturates. Roughly $90\%$ of trajectory pairs follow this latent-then-jump pattern, with the remainder exhibiting outlier profiles (Fig.~\ref{fig:cos_dist_lyapunov_outliers}).

However, we do not observe a smooth increase of the distance but
instead it is always characterized by large jumps. This phenomenology
is consistent with the standard picture of sensitivity to initial
conditions in discrete-space chaotic systems. In such systems, nearby
trajectories typically remain close while evolving within locally
weakly expanding regions of phase space, but separate rapidly once
they encounter directions associated with positive local Lyapunov
exponents. The observed jump-like divergence in hidden-state
space therefore provides qualitative evidence for strong sensitivity
to initial conditions, even though the discreteness of the token space
prevents a robust global estimate of the maximal Lyapunov exponent.

When the same analysis is performed using sentence-embedding
trajectories constructed from the generated text, the qualitative
two-regime structure remains, but the transition to large separation
becomes noticeably smoother as shown in
Fig.~\ref{fig:distance_embed}. Rather than abrupt bursts, the distance
curves exhibit a more gradual growth phase prior to saturation.
This is due to the implicit coarse-graining of the
sentence-level embedding. This has important implications as it shows
that the LLMs do not encode the distant future in the actual
tokens but rather that is discovered on the fly, which in the
end, may result in a completely different path due to small noise in
the initial conditions.

The geometry of the embedding space offers a natural interpretation of the observed two-regime structure. LLM sentence embeddings occupy a highly nonuniform manifold: most of the ambient space is sparsely populated, while model outputs concentrate in semantically meaningful clusters~\cite{ethayarajh2019contextual,rudman2023stable}. As a result, trajectories spend extended periods within a local semantic basin, producing low-amplitude fluctuations in pairwise distance. Intermittently, the dynamics carry the system across basin boundaries; because these transitions traverse relatively empty regions of embedding space, small token-level perturbations can project onto strongly expanding directions, producing the large separation events observed in the divergence curves.

The consistency of this behavior across distance metrics indicates that it reflects intrinsic properties of the underlying high-dimensional dynamics rather than artifacts of a particular distance definition. Different metrics nevertheless weight the separation process differently-some emphasizing onset time and others extremal deviation-highlighting the utility of complementary probes when estimating Lyapunov growth in discrete, high-dimensional generative systems.

\begin{figure}[h]
    \centering
    \begin{subfigure}[t]{0.225\textwidth}
        \includegraphics[width=\linewidth]{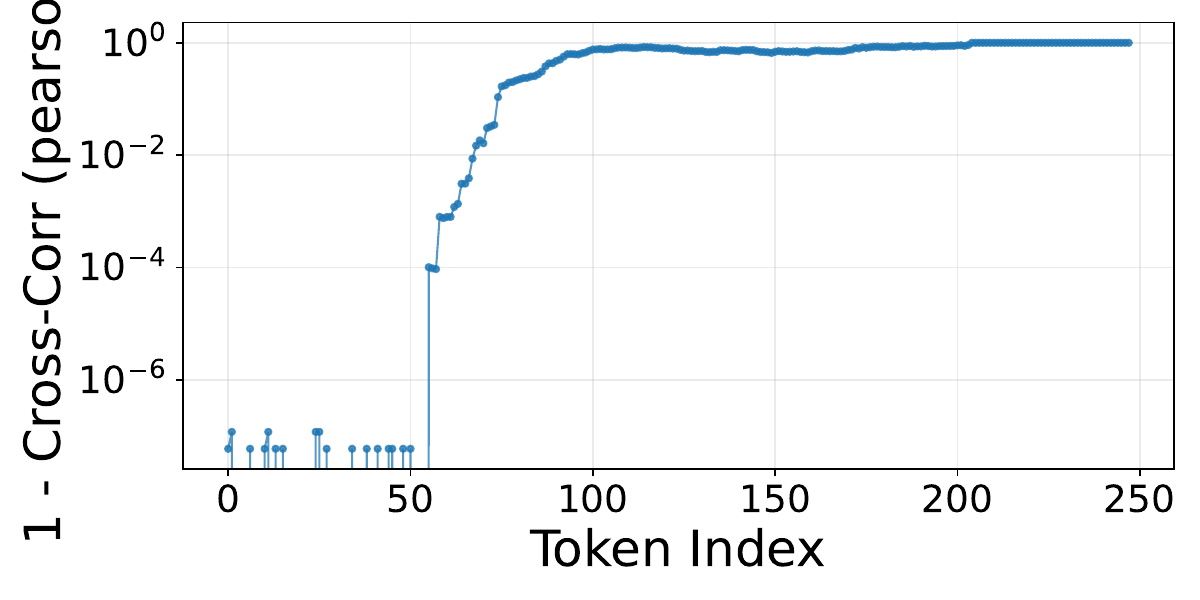}
        \caption{Cross-correlation dissimilarity}
    \end{subfigure}\hfill
    \begin{subfigure}[t]{0.225\textwidth}
        \includegraphics[width=\linewidth]{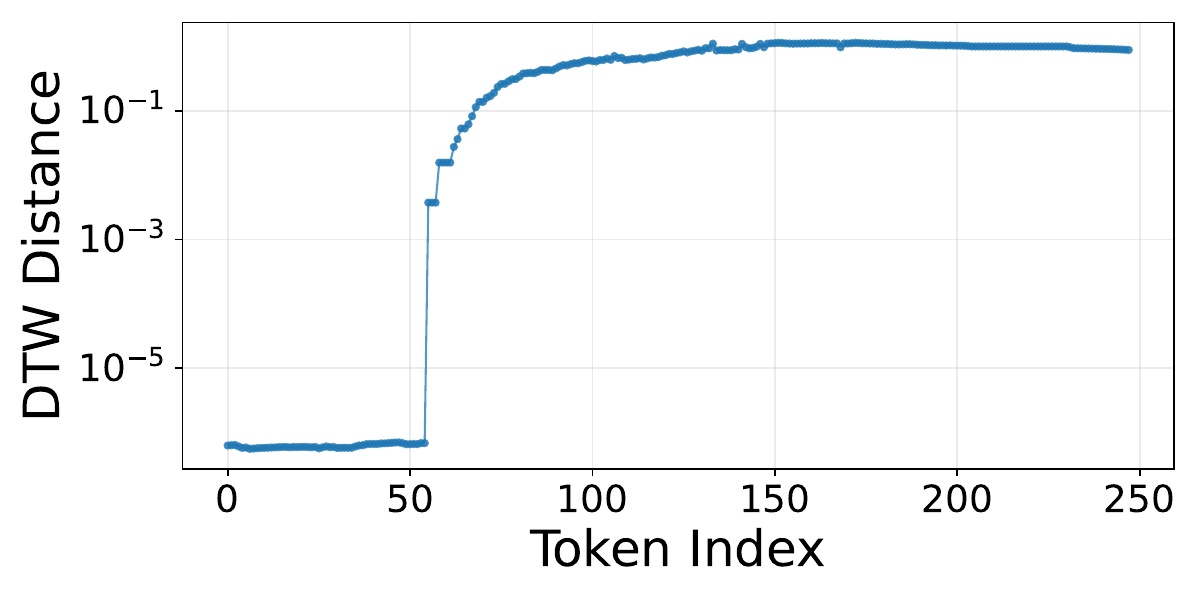}
        \caption{DTW}
    \end{subfigure}
    \\[0.5em]
    \begin{subfigure}[t]{0.225\textwidth}
        \includegraphics[width=\linewidth]{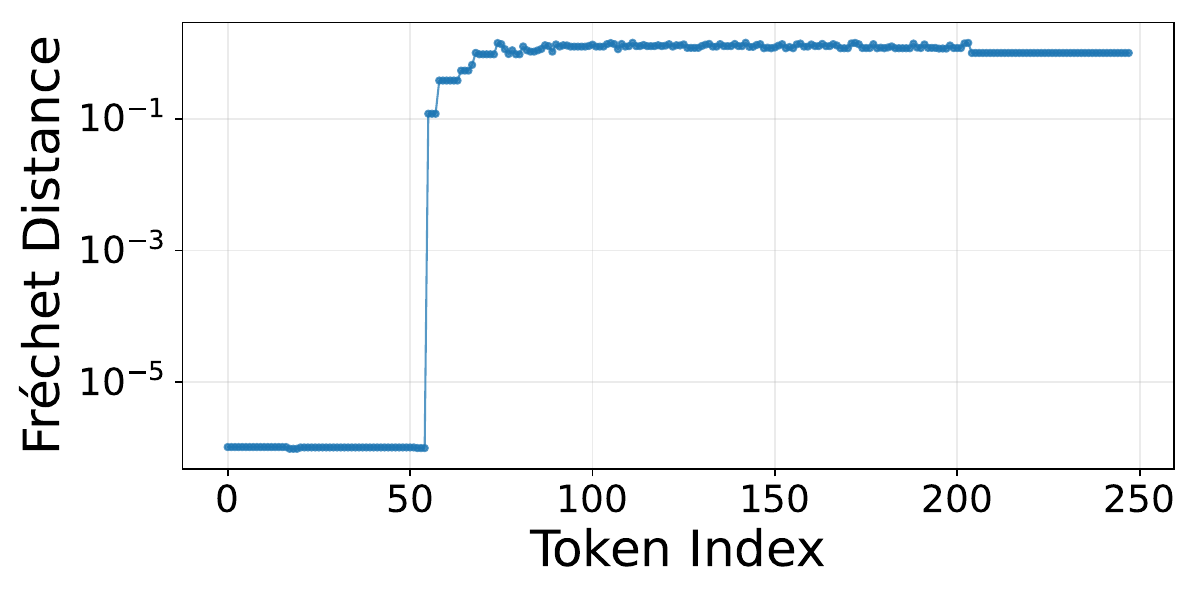}
        \caption{Fréchet}
    \end{subfigure}\hfill
    \begin{subfigure}[t]{0.225\textwidth}
        \includegraphics[width=\linewidth]{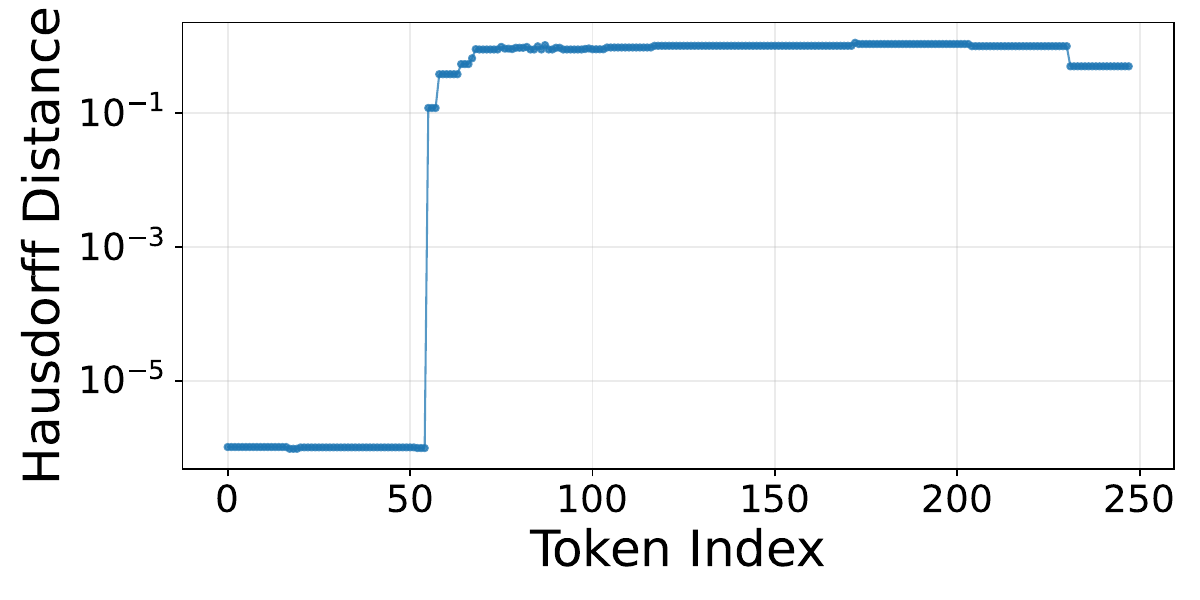}
        \caption{Hausdorff}
    \end{subfigure}
    \caption{Divergence between a pair of hidden-state trajectories measured using different distance metrics. Some metrics exhibit gradual divergence (panels (b), (d)), while others show abrupt, step-like separation (panels (a), (c)), corresponding to the first differing token.}
    \label{fig:distance_hidden}
\end{figure}

\begin{figure}[h]
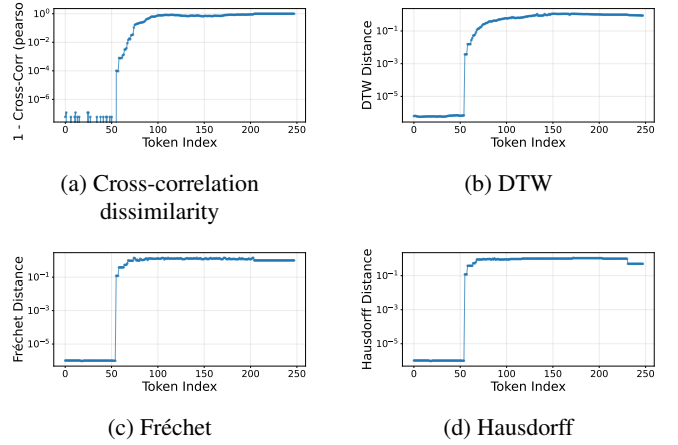

    \centering
    \begin{subfigure}[t]{0.225\textwidth}
        \includegraphics[width=\linewidth]{cross_corr_pearson_0_2_window_size_16.pdf}
        \caption{Cross-correlation dissimilarity}
    \end{subfigure}\hfill
    \begin{subfigure}[t]{0.225\textwidth}
        \includegraphics[width=\linewidth]{dtw_timeseries_0_2_window_size_16.pdf}
        \caption{DTW}
    \end{subfigure}
    \\[0.5em]
    \begin{subfigure}[t]{0.225\textwidth}
        \includegraphics[width=\linewidth]{frechet_timeseries_0_2_window_size_16.pdf}
        \caption{Fréchet}
    \end{subfigure}\hfill
    \begin{subfigure}[t]{0.225\textwidth}
        \includegraphics[width=\linewidth]{hausdorff_timeseries_0_2_window_size_16.pdf}
        \caption{Hausdorff}
    \end{subfigure}
    \caption{Divergence between sentence embedding trajectories measured using different distance metrics. In the case of cross correlation distance, the fluctuation increases dramatically, while for other metrics, we experience a normalization effect (notice that the scale of the Y axis is much smaller on panels (b), (c), and (d) than on Fig~\ref{fig:distance_hidden}).}
    \label{fig:distance_embed}
\end{figure}

\subsection{Dependence on the Attention Window}
\label{subsec:attn_window}

The latent-then-jump phenomenology above implies that a perturbation must be carried forward in time before it can flip a token. Within the Transformer architecture, self-attention is the only mechanism that propagates information, and hence perturbations, from earlier to later positions in the sequence. To test the causal role of attention directly, we restrict the attention span to a sliding window of $w$ tokens and measure how the first-divergence time depends on $w$.

\begin{figure}[H]
\centering
\includegraphics[width=\linewidth]{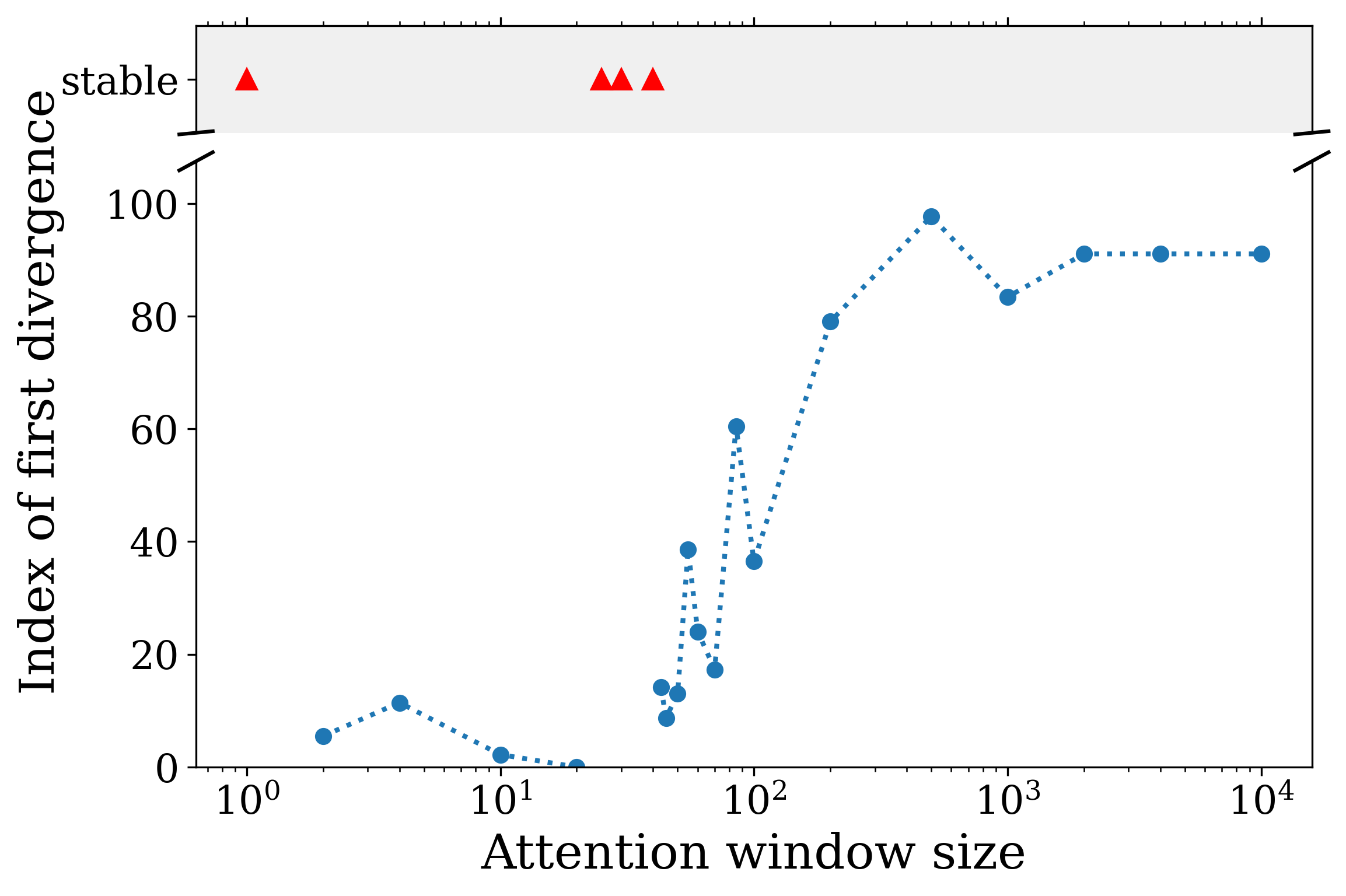}
\caption{Dependence of the divergence on the attention window size $w$ (perturbation magnitude $10^{-4}$). The vertical axis shows the number of generated tokens until a pair of trajectories first selects a different token; ensembles that never diverge within the generation horizon are marked as ``stable''.}
\label{fig:attn_window}
\end{figure}

Figure~\ref{fig:attn_window} reveals a nontrivial, non-monotonic dependence. For $w=1$, each token attends only to itself, no perturbation can propagate, and the dynamics are trivially stable. For very small windows, the model is dominated by its immediately preceding (perturbed) tokens and diverges almost instantly. For intermediate window sizes, an interesting stabilization occurs: the large shared context suppresses the effect of the perturbation on the first generated tokens, and once identically generated tokens fill the window, the perturbed prompt slides out of the attention span entirely, so the trajectories never diverge. Only for sufficiently large windows, where the perturbed prompt remains within the attention span, do we recover consistent but delayed divergence. This non-monotonic behavior, absent from any token-wise mechanism, identifies the attention coupling as the carrier of chaotic perturbation growth, complementing the operator-level result of Sec.~\ref{subsec:layer_perturbations} that the attention Jacobian is expansive. The precise shape of the curve depends on the perturbation magnitude; qualitatively similar behavior was found for the second model studied in the underlying thesis~\cite{jaca2026thesis}.

\subsection{Recurrence Plots}
\label{subsec:rp_results}

Recurrence plots (RPs) provide a geometric visualization of hidden-state trajectories and reveal complex temporal structure reminiscent of classical chaotic systems such as the Lorenz attractor~\cite{DeterministicNonperiodicFlow} (see Fig.~\ref{fig:rp_examples}). To examine how this structure develops across the network, we constructed recurrence plots both for the representations after the first and last layers of the model.

\begin{figure}[h]
    \centering
    \begin{subfigure}[t]{0.225\textwidth}
        \includegraphics[width=\linewidth]{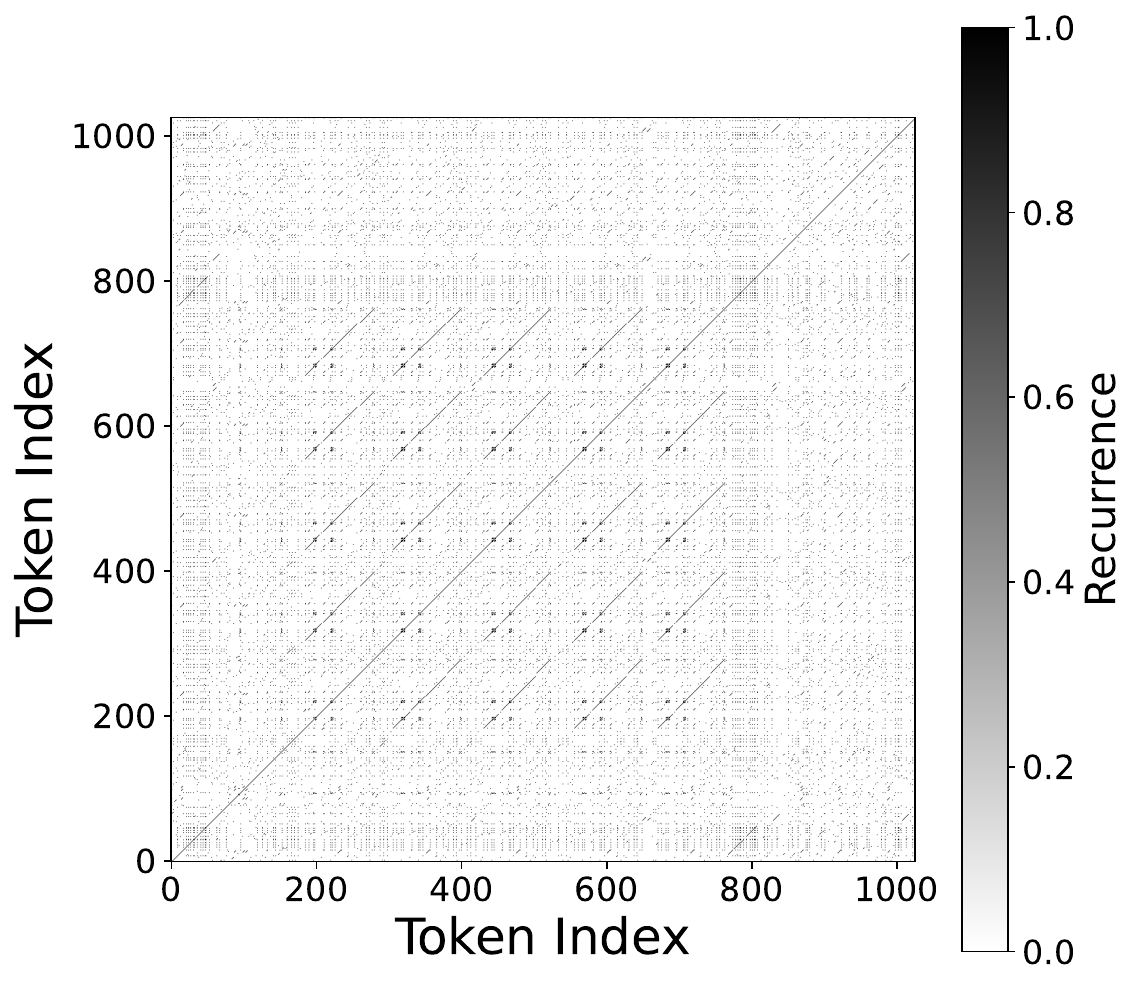}
        \caption{Prompt 1, first layer}
    \end{subfigure}\hfill
    \begin{subfigure}[t]{0.225\textwidth}
        \includegraphics[width=\linewidth]{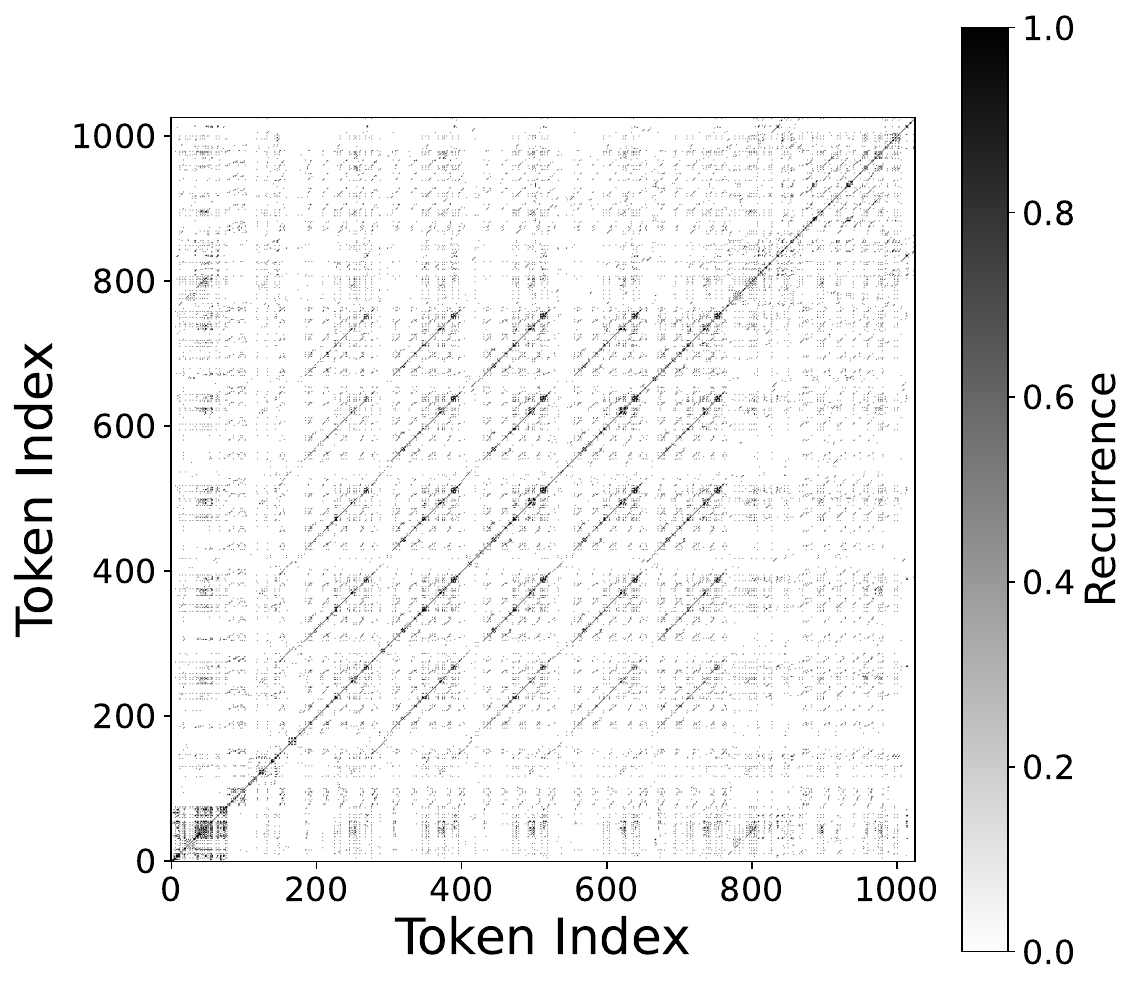}
        \caption{Prompt 1, last layer}
    \end{subfigure}
    \\[0.5em]
    \begin{subfigure}[t]{0.225\textwidth}
        \includegraphics[width=\linewidth]{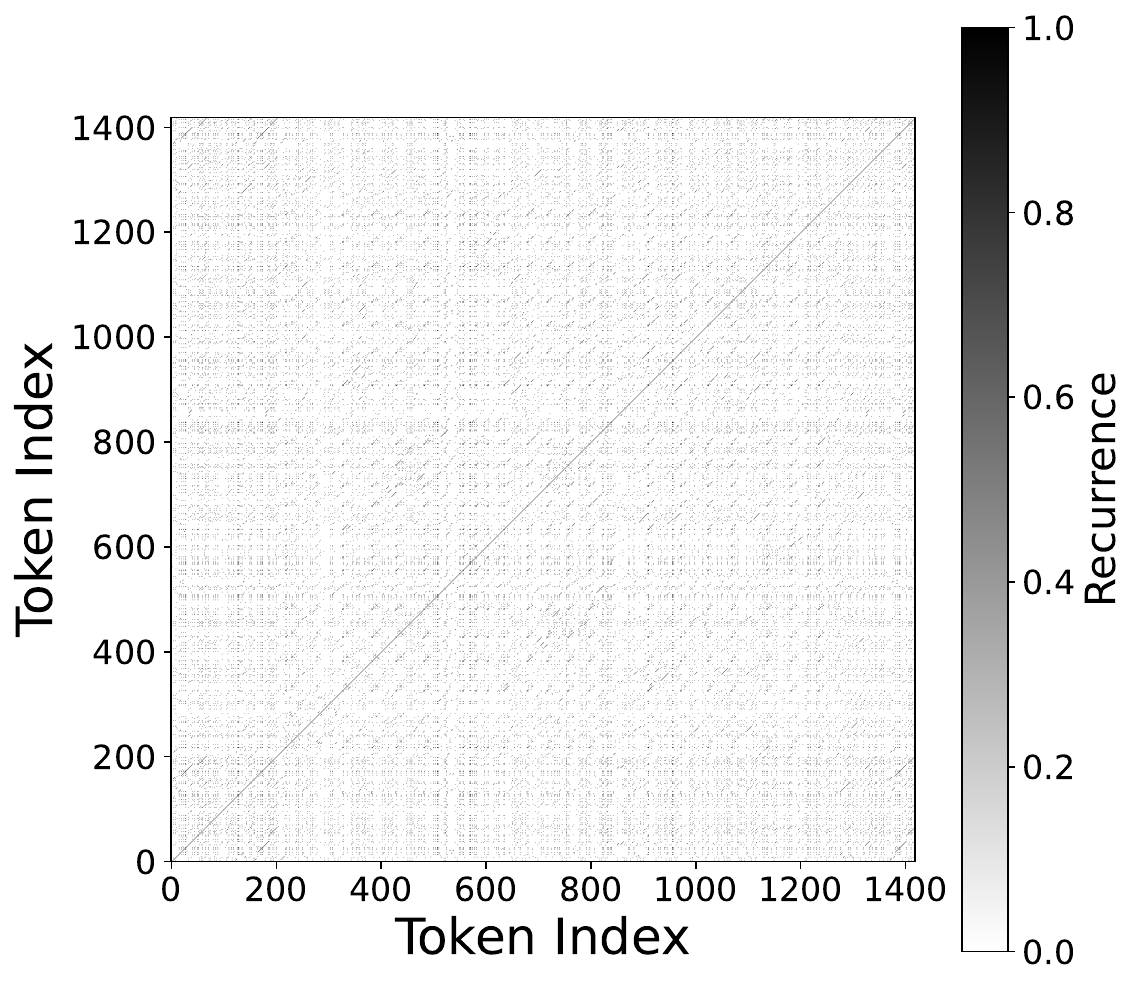}
        \caption{Prompt 2, first layer}
    \end{subfigure}\hfill
    \begin{subfigure}[t]{0.225\textwidth}
        \includegraphics[width=\linewidth]{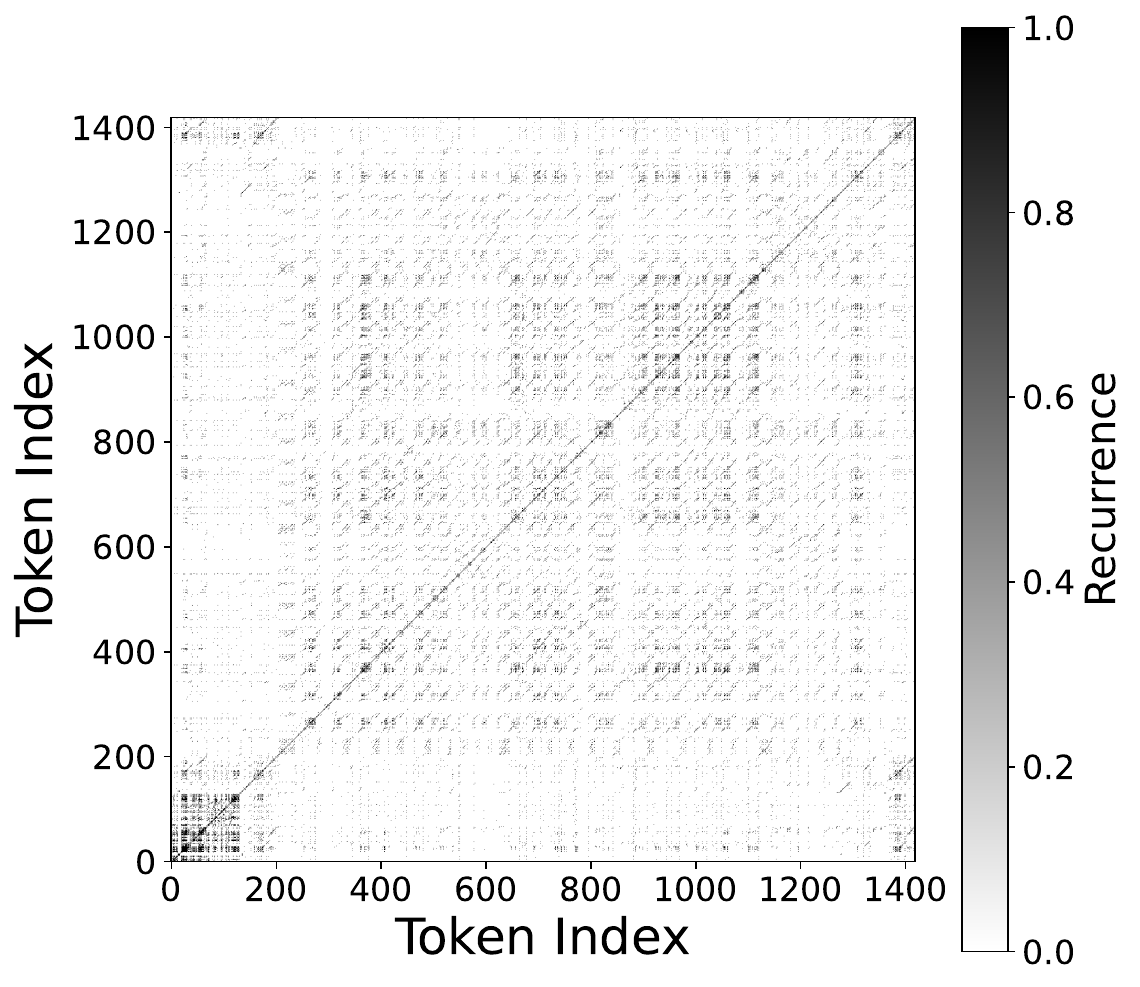}
        \caption{Prompt 2, last layer}
    \end{subfigure}
    \caption{Recurrence plots for different prompts and transformer layers using cosine distance. Chaotic features such as diagonal line structures become more pronounced in deeper layers (cf. panel (b)). The recurrence rate was fixed to $0.03$.}
    \label{fig:rp_comparison_layers}
\end{figure}

Interestingly, the recurrence plots of the first-layer representations display patterns closer to those expected from stochastic signals. The plots contain relatively few coherent diagonal structures and consist largely of isolated recurrence points. Such patterns indicate that neighboring trajectory segments do not remain similar for extended periods, suggesting dynamics dominated by local fluctuations rather than deterministic structure. This behavior is consistent with the role of early layers, which primarily encode token-level embeddings and local lexical information. As different tokens at this level have only their base representation vector, words put after each other look more like a random walk. In extremely high dimensions this means uncorrelated large distance.

In contrast, recurrence plots of deeper layers reveal substantially richer geometric organization. As the words obtain components related to the meaning of the text they get more and more similar for prolonged amount of time. A prominent feature across prompts and layers is thus the presence of numerous short diagonal line segments with varying lengths. In recurrence analysis, such fragmented diagonals indicate that nearby trajectory segments remain similar only for finite intervals before separating, a signature consistent with sensitivity to initial conditions and positive Lyapunov exponents. The abundance of diagonals with a broad length distribution is therefore compatible with the chaotic dynamics suggested by the divergence analysis.

In some cases the recurrence plots also display quasi-periodic stripe or grid-like patterns. Inspection of the corresponding outputs shows that these structures arise when the model generates strongly regular textual formats, such as enumerated suggestions or repeated recommendation templates. These patterns produce recurrent passages through similar regions of hidden-state space, giving rise to the periodic structures visible in the recurrence matrix.

Another robust feature observed across trajectories is the presence of block-like structures along the main diagonal. These blocks correspond to intervals during which the trajectory remains confined to a relatively localized region of state space before abruptly transitioning to a different region. In the context of language generation, such transitions frequently coincide with shifts in the semantic focus of the generated text.
\begin{figure}[h]
    \centering
    \begin{subfigure}[t]{0.225\textwidth}
        \centering
        \includegraphics[width=\linewidth]{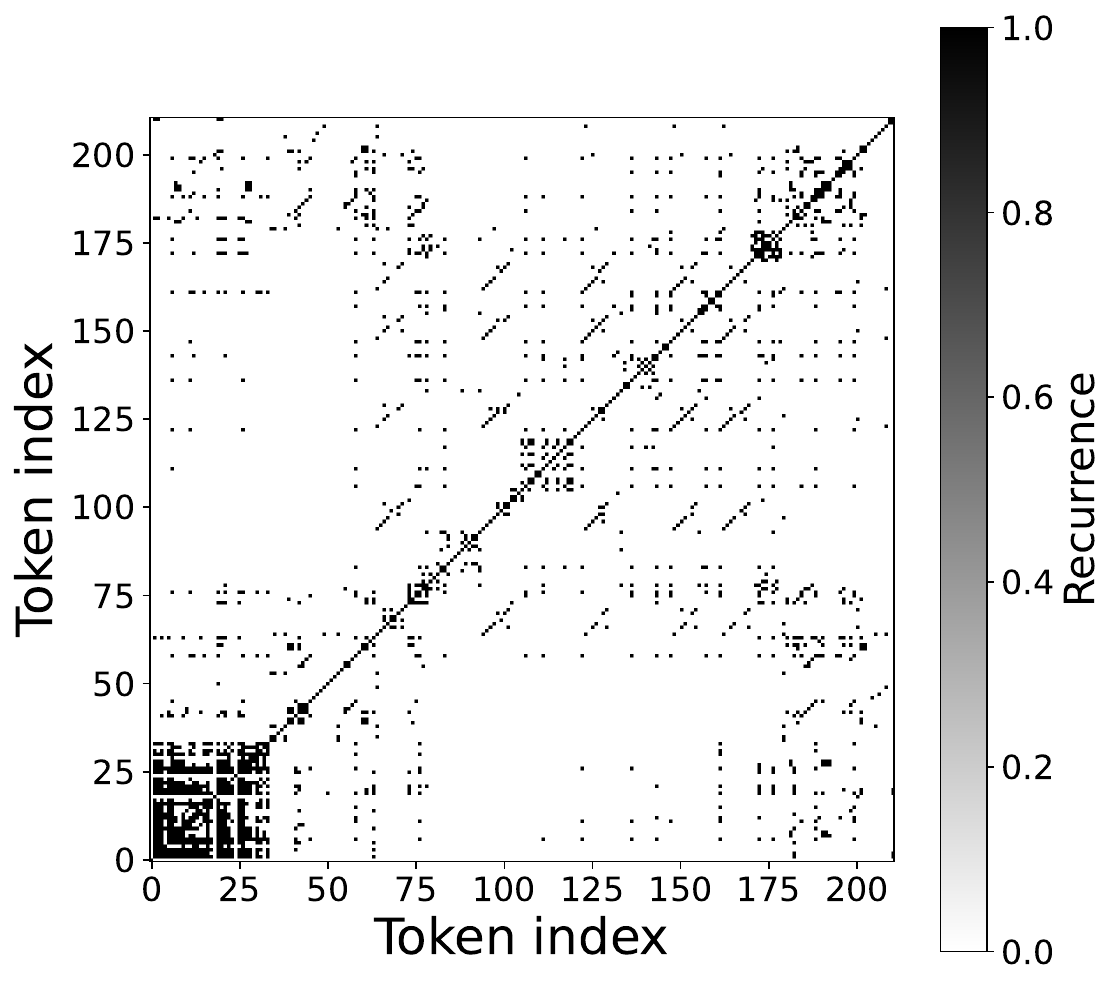}
        \caption{Prompt a}
    \end{subfigure}\hfill
    \begin{subfigure}[t]{0.225\textwidth}
        \centering
        \includegraphics[width=\linewidth]{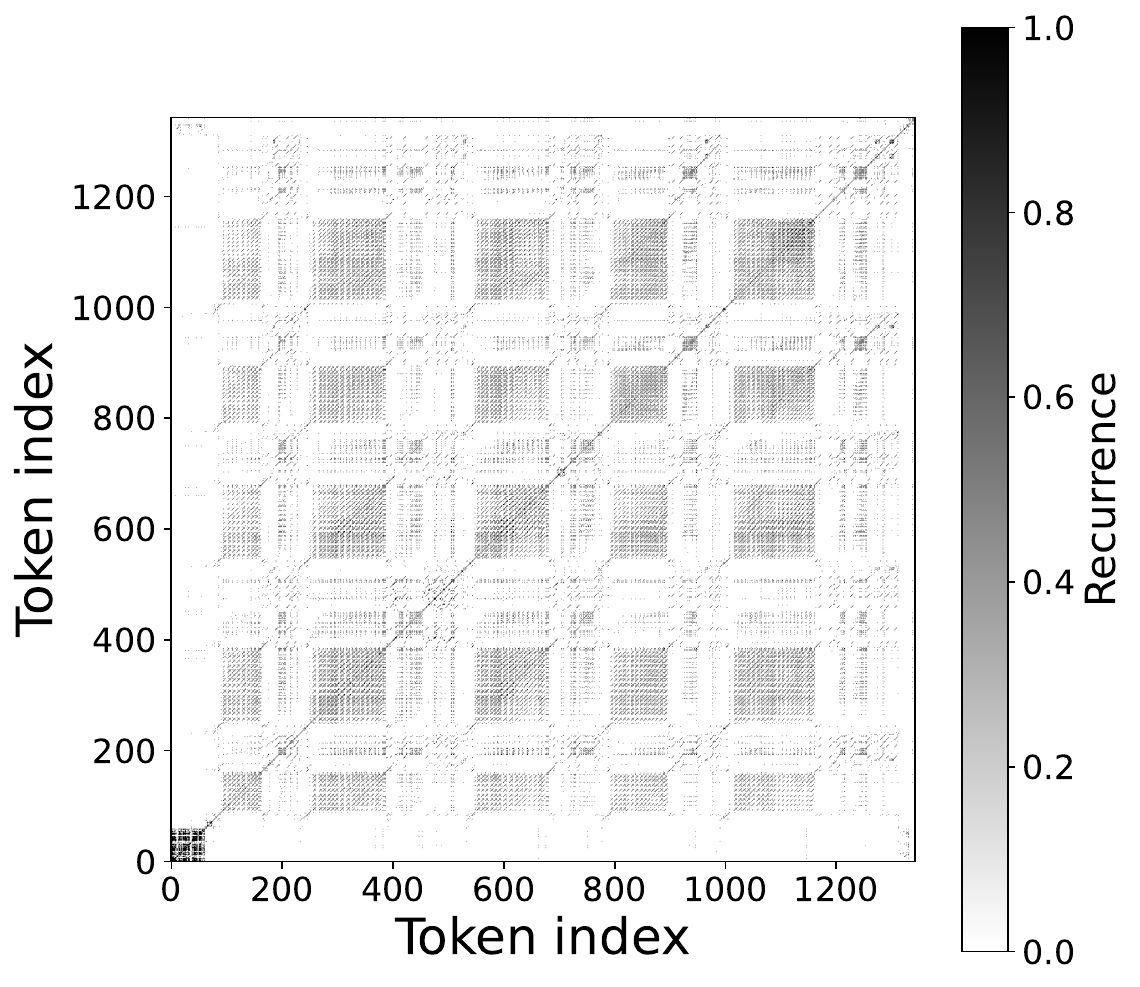}
        \caption{Prompt b}
    \end{subfigure}
    \\[0.5em]
    \begin{subfigure}[t]{0.225\textwidth}
        \centering
        \includegraphics[width=\linewidth]{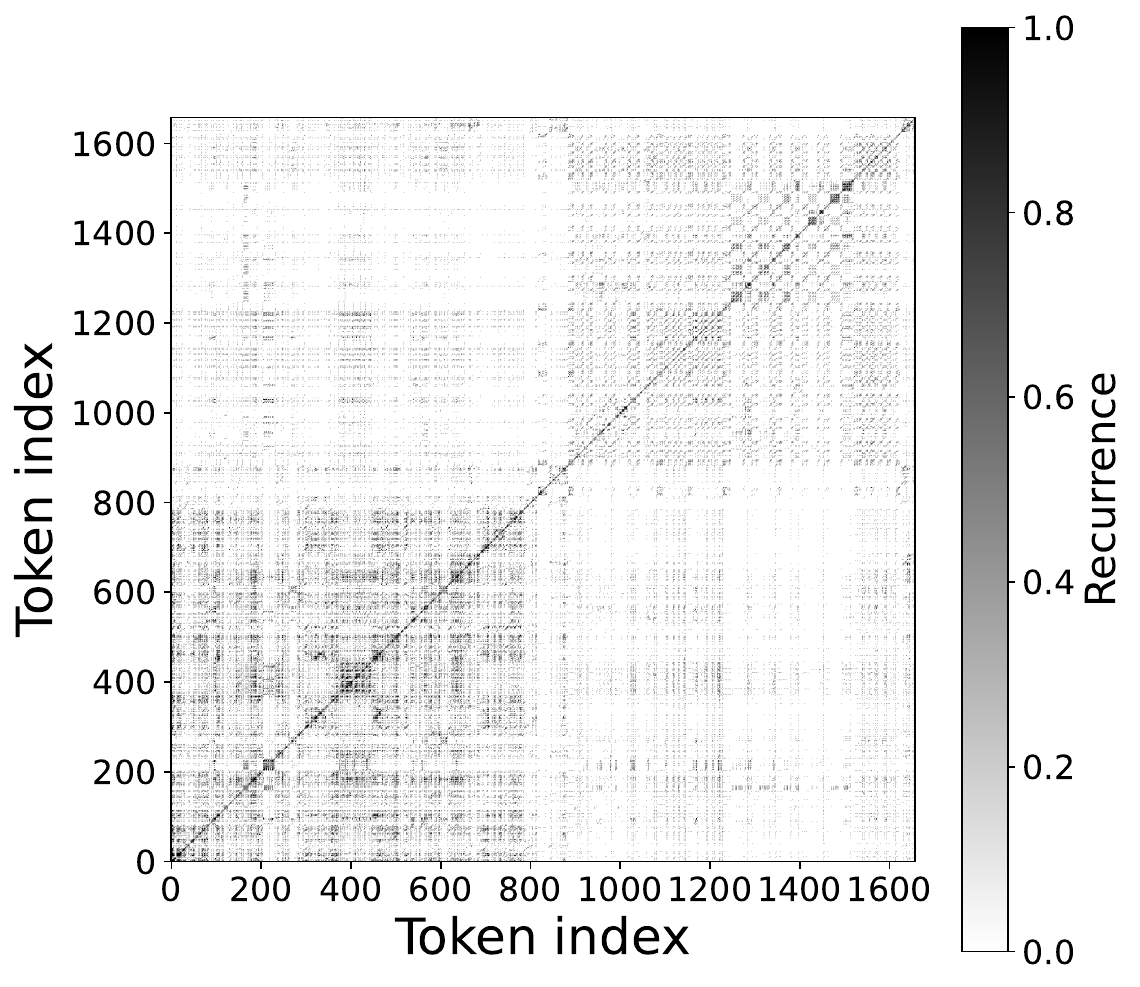}
        \caption{Prompt c}
    \end{subfigure}\hfill
    \begin{subfigure}[t]{0.225\textwidth}
        \centering
        \includegraphics[width=\linewidth]{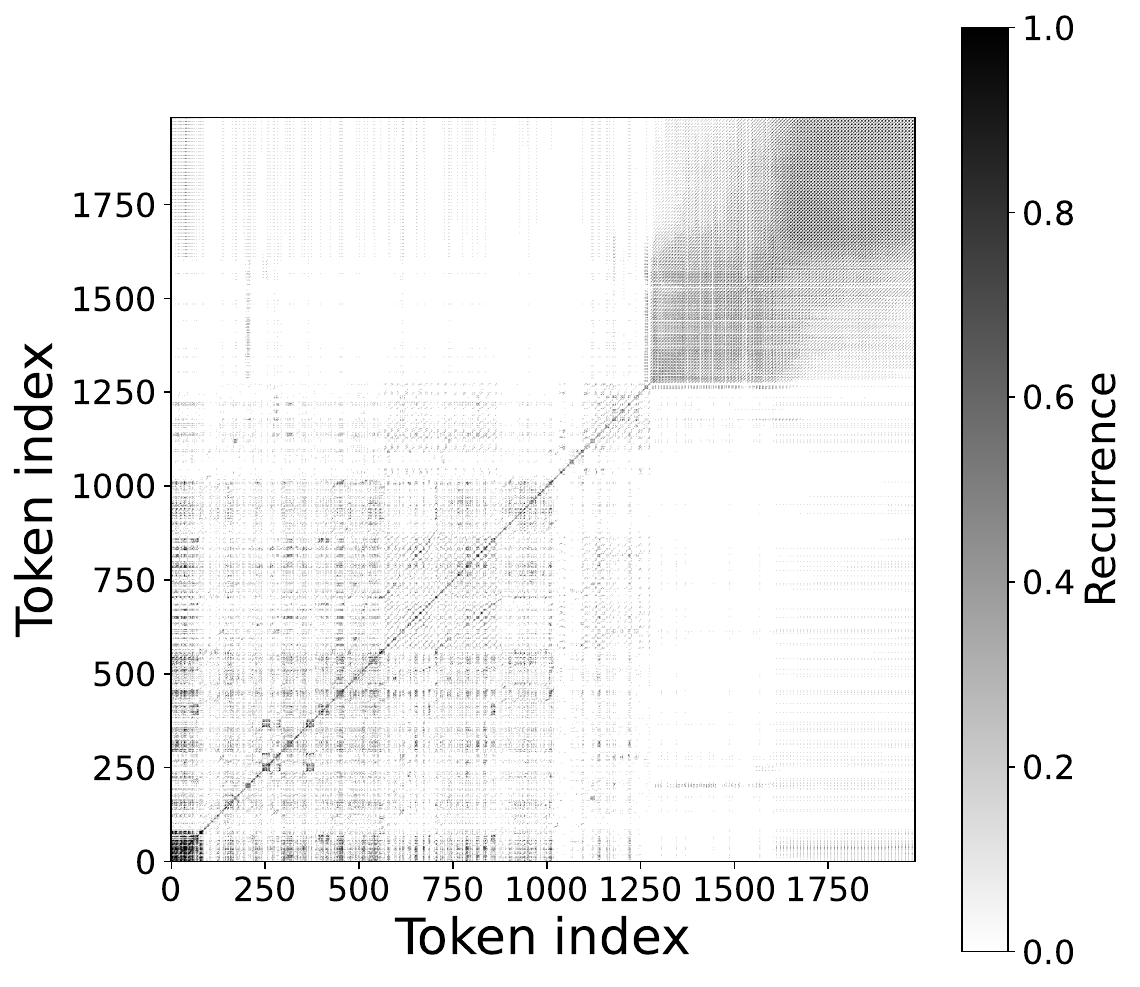}
        \caption{Prompt d}
    \end{subfigure}
    \caption{Recurrence plots obtained for different prompts, evaluated at the last layer. Panel (a) exhibits fractal-like behavior, with short diagonal lines, which can be interpreted as a state at the edge of chaos. The recurrence rate was fixed to $0.05$.}
    \label{fig:rp_interesting}
\end{figure}


\subsection{Dimensionality}
\label{subsec:dimensionality}

\begin{figure*}
    \centering

    \begin{subfigure}[t]{0.3\linewidth}
        \centering
        \includegraphics[width=\linewidth]{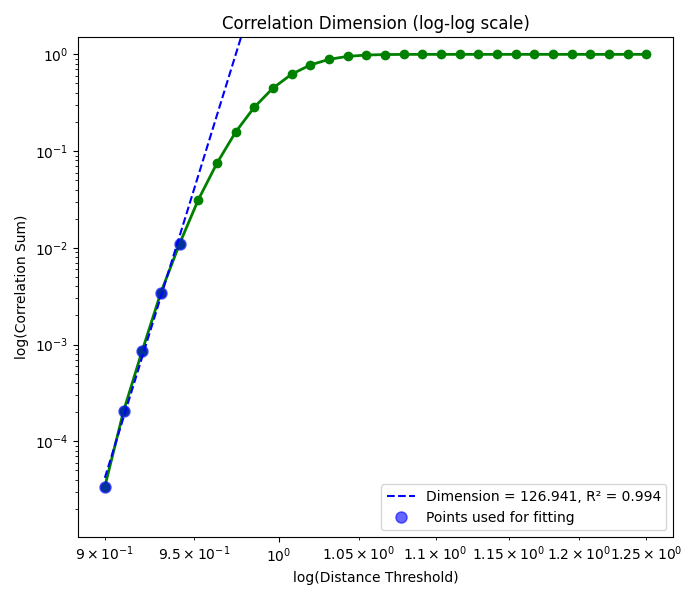}
        \caption{Random}
    \end{subfigure}\hfill
    \begin{subfigure}[t]{0.3\linewidth}
        \centering
        \includegraphics[width=\linewidth]{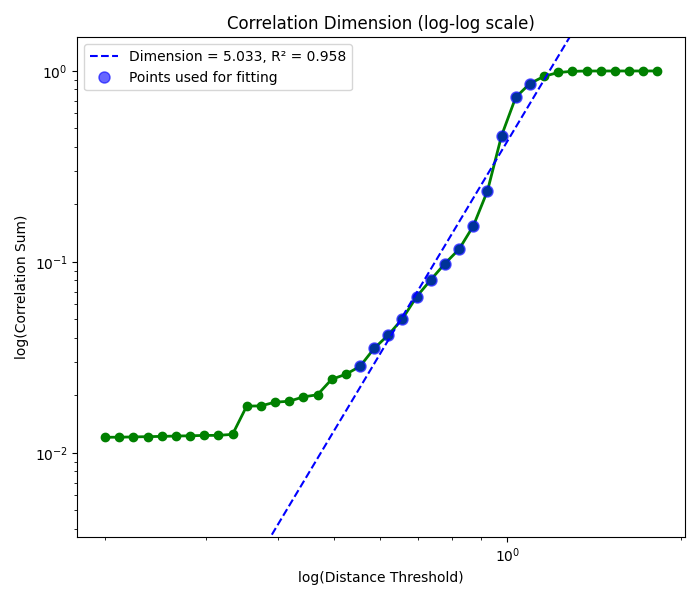}
        \caption{First layer}
    \end{subfigure}\hfill
    \begin{subfigure}[t]{0.3\linewidth}
        \centering
        \includegraphics[width=\linewidth]{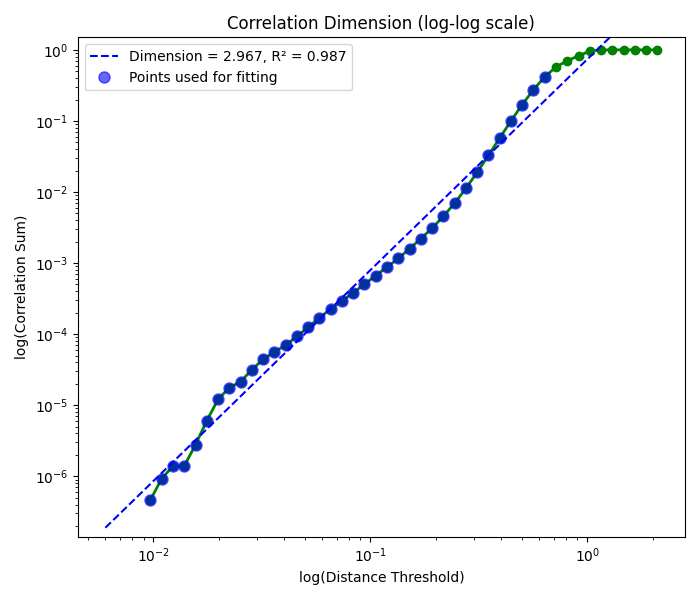}
        \caption{Last layer}
    \end{subfigure}


    \caption{Correlation dimension estimates for a random vector set and for first- and last-layer hidden-state trajectories.}
    \label{fig:dim_first_last}
\end{figure*}

Finally, we investigate the effective dimensionality of hidden-state trajectories using correlation dimension estimates. Figure~\ref{fig:dim_first_last} compares random vectors with trajectories from the first and last layers. While first-layer representations closely resemble random vectors---apart from repeated-token effects---the last-layer trajectories exhibit scaling behavior consistent with low-dimensional deterministic chaos.

The random-like geometry of the first layer is expected in high dimensions: the number of nearly orthogonal directions grows exponentially with the ambient dimension~\cite{cai2013distributionsanglesrandompacking}, a property that allows LLMs to superpose far more concepts than they have dimensions~\cite{blessing,park2024linearrepresentationhypothesisgeometry}. First-layer states are still close to the raw token embeddings, so distinct tokens produce nearly orthogonal vectors while repeated tokens produce exactly coincident points; only after contextual information accumulates through the layers do the trajectories organize onto a structured manifold. We emphasize that the low-dimensional structure detected at the last layer lives \textbf{within} the very high-dimensional ($D=1536$) hidden-state space: the effective trajectory dimension is orders of magnitude below the ambient dimension.

While the extracted correlation dimension depends on the choice of distance metric, the qualitative behavior of the correlation sum is robust across metrics. We  therefore interpret the correlation dimension not as a precise estimate of the fractal dimension of the attractor, but as an indicator of deterministic structure and effective dimensionality.

\begin{figure}[h]
    \centering
    \begin{subfigure}[t]{0.225\textwidth}
        \includegraphics[width=\linewidth]{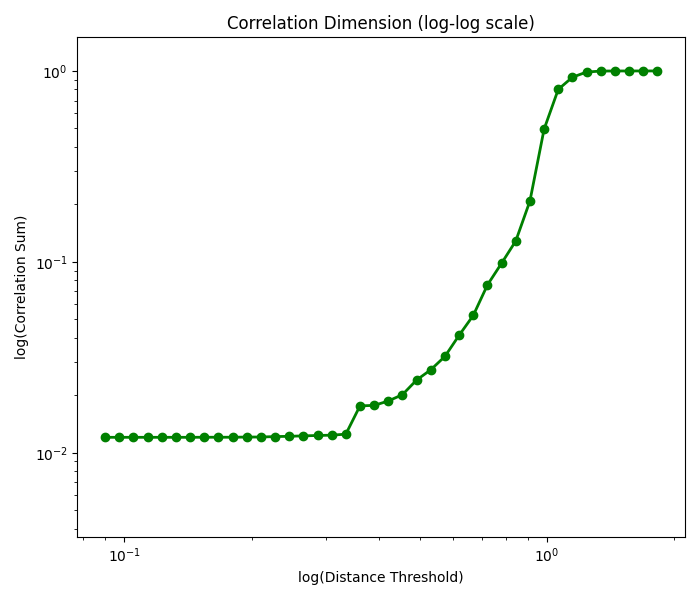}
        \caption{First layer}
    \end{subfigure}\hfill
    \begin{subfigure}[t]{0.225\textwidth}
        \includegraphics[width=\linewidth]{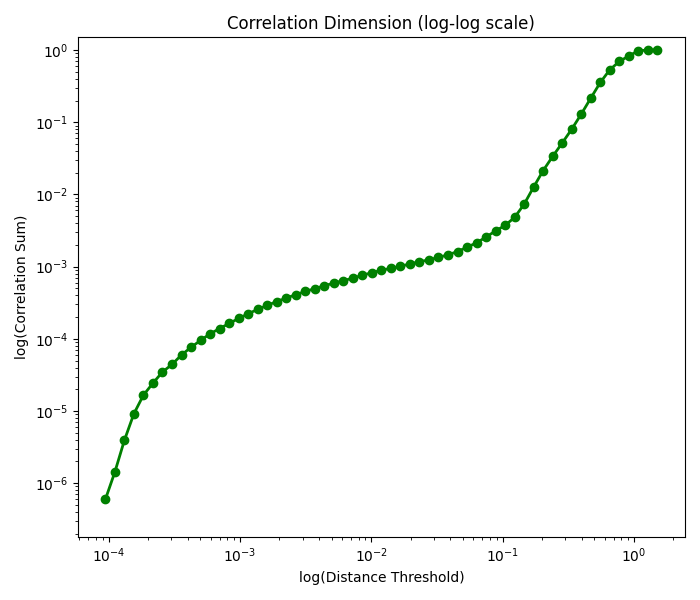}
        \caption{Last layer}
    \end{subfigure}
    \caption{Correlation dimension plots under probabilistic sampling ($T=0.6$).}
    \label{fig:dim_first_last_temperature}
\end{figure}

Under probabilistic sampling (Fig.~\ref{fig:dim_first_last_temperature}), the scaling structure observed in deterministic generation is disrupted at the last layer, while the first layer remains largely unaffected. This further supports the conclusion that chaotic dynamics primarily emerge in deeper layers and are suppressed when stochasticity dominates token selection.

Overall, these results indicate that early layers of the model behave similarly to high-dimensional random embeddings, while deeper layers organize trajectories onto structured, low-dimensional manifolds exhibiting sensitivity to initial conditions. This layer-wise transition is consistent with the emergence of deterministic chaotic dynamics in the final stages of text generation.

\section{Conclusions and Discussion}
\label{sec:conclusions}

In this work, we investigated the internal dynamics of a large language model through the framework of nonlinear dynamics and chaos theory. By interpreting autoregressive generation as a high-dimensional discrete-time dynamical system, analyzing trajectories in both hidden-state and semantic embedding spaces via sentence embedders, and complementing these global diagnostics with an exact Jacobian analysis of the model's architectural sub-blocks, we sought to determine whether the evolution of representations within Transformer-based LLMs with CoT exhibits signatures of deterministic chaos.

Our results provide consistent evidence for the possibility of such behavior. Across multiple distance metrics-capturing fundamentally different geometric notions such as alignment, extremal deviation, temporal warping, and global self-similarity and across different representational spaces, we observed qualitatively robust patterns of trajectory divergence, recurrence, and boundedness. This invariance with respect to metric choice and representation strongly suggests that the observed phenomena reflect intrinsic properties of the internal dynamics of the model, rather than artifacts of a particular analytical tool.

A central finding of this study is that the effective dynamical state of an LLM cannot be adequately captured by a single hidden-state vector. Although the next-token prediction depends explicitly on the last-layer hidden state of the current token, our analysis shows that distances computed on sequences of individual hidden states are often noisy and poorly correlated with semantic divergence. This motivated the use of sliding-window representations and sentence-level embeddings, which implicitly incorporate information from a finite history of tokens. From a dynamical systems perspective, this places LLMs closer to the class of \textbf{delay systems}, in which state evolution depends on a window of past states rather than on an instantaneous configuration. The success of these approaches reflects the architectural reality of Transformers, whose transition function depends on the entire context window via self-attention.

Notably, divergence patterns extracted from hidden-state trajectories and from sentence embedding trajectories were strikingly similar. This is not a trivial observation: LLM hidden states are not explicitly trained to form a semantically interpretable metric space, whereas sentence embedding models are optimized precisely for this purpose. The agreement between these two representations indicates that, at least in later layers, distances in hidden-state space encode meaningful semantic and dynamical information. This alignment further supports the view that chaotic-like behavior is a genuine property of the model’s internal representation dynamics.

From a nonlinear dynamics standpoint, our findings support an interpretation of LLMs as high-dimensional, nonlinear systems operating near the \textbf{edge of chaos}. We observe strong sensitivity to initial conditions, manifested as intermittent, jump-like divergence of nearby trajectories once a latent perturbation crosses the token-sampling threshold, with a broad distribution of divergence onset times; the dependence of this divergence on the attention window size identifies self-attention as the mechanism carrying the perturbation forward; structured but non-periodic recurrence plots reminiscent of canonical chaotic systems; and dimensionality estimates consistent with trajectories evolving on a lower-dimensional, fractal-like manifold rather than exploring the full embedding space. Architecturally, this regime naturally arises from the interplay between components that promote expansion and coupling-most notably self-attention and nonlinear feedforward layers-and components that enforce boundedness and contraction, such as normalization layers. The resulting balance between divergence and confinement is a hallmark of chaotic attractors.

This qualitative picture is corroborated quantitatively by the operator-level Jacobian analysis of Sec.~\ref{subsec:layer_perturbations}: the static weights and measured Jacobians of both self-attention and the feed-forward block exceed unit spectral norm at essentially every layer, whereas the normalization Jacobian acts as a scaled orthogonal projection that annihilates the radial, most-expansive component of a perturbation before it reaches the next nonlinear sub-block, with the residual connection carrying the surviving perturbation forward. A closed-form mean-field approximation for the feed-forward expansion factor, derived from this same Jacobian, agrees closely with the exact measurement across all layers, giving a predictive, weight-only diagnostic for a component's contribution to chaotic stretching. The resulting layer-wise profile, expansion through the middle of the network followed by contraction in the last few layers, echoes the non-normal-to-symmetric spectral transition and low-rank bottleneck recently reported from direct Jacobian eigendecomposition of production-scale LLMs~\cite{fernando2026transformer_residual_stream_dynamics}, suggesting that this stretch-then-fold organization of the residual stream may be a general feature of trained Transformer language models rather than an artifact of the specific model studied here.

Such a dynamical regime may help explain the dual capability of LLMs to maintain long-range coherence while remaining highly flexible and responsive to small input variations. Systems operating near the edge of chaos are known to maximize information propagation~\cite{info_edge_chaos}, \cite{LANGTON199012}, computational expressivity, and sensitivity without losing global stability, properties that resonate strongly with the observed behavior of modern language models.

At the same time, this study highlights the challenges of applying classical chaos analysis techniques in extremely high-dimensional, discrete, and partially stochastic settings. Concepts such as Lyapunov spectra and fractal dimension must be interpreted with care, and their estimation requires methodological adaptations. Rather than treating these quantities as precise invariants, we emphasize their value as comparative and diagnostic tools for probing structure, sensitivity, and organization across layers, decoding regimes, and representations.

Our trajectory-level analysis was limited to a single distilled decoder-only model, although the operator-level Jacobian and perturbation results were reproduced on a second, architecturally distinct model (\texttt{Phi-4-mini-instruct}). While we expect many of the observed phenomena to generalize across architectures and scales, architectural features-such as Chain-of-Thought prompting, attention patterns, normalization schemes, and context length-clearly leave identifiable signatures in the dynamics, for instance, in recurrence plots. Future work should therefore systematically investigate how chaotic characteristics vary with model size, training regime, and architectural choices. Comparing encoder-only, decoder-only, and encoder–decoder models may clarify the role of causal versus bidirectional attention, while varying context length and normalization strategies could reveal which components most strongly regulate dynamical stability.

More broadly, our results suggest that tools from nonlinear dynamics provide a promising and largely unexplored framework for understanding the internal behavior of large language models. By bridging modern machine learning architectures with the theory of complex systems, this perspective opens new avenues for interpretability, robustness analysis, and the study of emergent computation in artificial neural networks.

\begin{acknowledgments}
Support from the Hungarian National Research, Development and Innovation Office NKFIH (149429, TKP2021-NVA-02 and TKP2021-EGA-02) is acknowledged.
Kristóf Benedek acknowledges the support from the Doctoral Excellence Fellowship Programme (DCEP), founded by the National Research Development and Innovation Fund of the Ministry of Culture and Innovation and the Budapest University of Technology and Economics. 
Results were obtained by using the HPC cluster of the Budapest University of Technology and Economics, Department of Theoretical Physics.
\end{acknowledgments}

\section*{Data Availability Statement}

The data that support the findings of this study are available from the corresponding author upon reasonable request.

\section*{References}
\bibliography{main}
\newpage

\appendix
\section{Mean-Field Derivation for the Feed-Forward Expansion}
\label{app:mft}

For a simplified feed-forward block $\text{MLP}(\mathbf{x})=W_2\phi(W_1\mathbf{x})$ mapping a single token $\mathbf{x}\in\mathbb{R}^D$ (a simpler surrogate for the SwiGLU block of Sec.~\ref{subsec:layer_perturbations}, used here only to keep the derivation transparent), the Jacobian is $\mathbf{J}=W_2\Sigma W_1$, with $\Sigma=\text{diag}(\phi'(W_1\mathbf{x}))\in\mathbb{R}^{D_{\text{ff}}\times D_{\text{ff}}}$. The isotropic expansion factor introduced in Sec.~\ref{subsec:layer_perturbations} is
\begin{equation}
\bar\lambda = \frac{1}{D}\|\mathbf{J}\|_F^2 = \frac{1}{D}\text{Tr}(\mathbf{J}\mathbf{J}^T).
\end{equation}
Using the cyclic property of the trace and defining the Gramian matrices $H_1=W_1W_1^T$, $H_2=W_2^TW_2$,
\begin{align}
\|\mathbf{J}\|_F^2 &= \text{Tr}(H_2\Sigma H_1\Sigma) = \sum_{i}(H_2)_{ii}(H_1)_{ii}(\phi_i')^2 + \notag \\
&\quad +\sum_{i\ne j}(H_2)_{ij}(H_1)_{ji}\phi_i'\phi_j'.
\end{align}
We make two mean-field simplifications, standard in the study of both random and trained feedforward networks~\cite{poole2016exponential_expressivity,schoenholz2017deep_information_propagation,yang2017meanfield_resnets_edge_of_chaos}: (i) the off-diagonal sum is neglected, since for $D_{\text{ff}}\gg1$ the cross-terms behave as zero-mean, alternating-sign random variables that destructively interfere; and (ii) the diagonal Gramian entries are replaced by their row/column-average value, $(H_k)_{ii}\approx \text{Tr}(H_k)/D_{\text{ff}}=\|W_k\|_F^2/D_{\text{ff}}$. Substituting into the diagonal sum gives
\begin{equation}
\|\mathbf{J}\|_F^2 \approx \frac{\|W_1\|_F^2\|W_2\|_F^2}{D_{\text{ff}}}\,\gamma, \qquad \gamma=\frac{1}{D_{\text{ff}}}\sum_i(\phi_i')^2,
\end{equation}
and, defining the scaled Frobenius norms $\tilde F_1,\tilde F_2$,
\begin{equation}
\bar\lambda \approx \tilde F_1\, \tilde F_2\, \gamma.
\end{equation}
For the SwiGLU block used in the model, the gating mechanism splits this three-term product into the two-term sum quoted in Sec.~\ref{subsec:layer_perturbations}, $\bar\lambda\approx \tilde F_{\text{down}}\left[\tilde F_{\text{gate}}\,\overline{d_x^2}+\tilde F_{\text{up}}\,\overline{s_x^2}\right]$, with $\gamma$ replaced by the measured activation densities $\overline{d_x^2},\overline{s_x^2}$.

We test both assumptions numerically: for (i), we compute $R=\lambda_{\text{diagonal}}/\lambda_{\text{true}}$, the fraction of the exact expansion accounted for by the diagonal term alone ($R\approx1$ indicates diagonal dominance); for (ii), we compute the coefficient of variation $\text{CV}=\sigma/\mu$ of the row/column norms of $W_{\text{gate}}$, $W_{\text{up}}$, $W_{\text{down}}$ ($\text{CV}\ll1$ indicates uniformity).

\begin{figure}[H]
\centering
\begin{subfigure}[b]{0.48\linewidth}
\centering
\includegraphics[width=\linewidth]{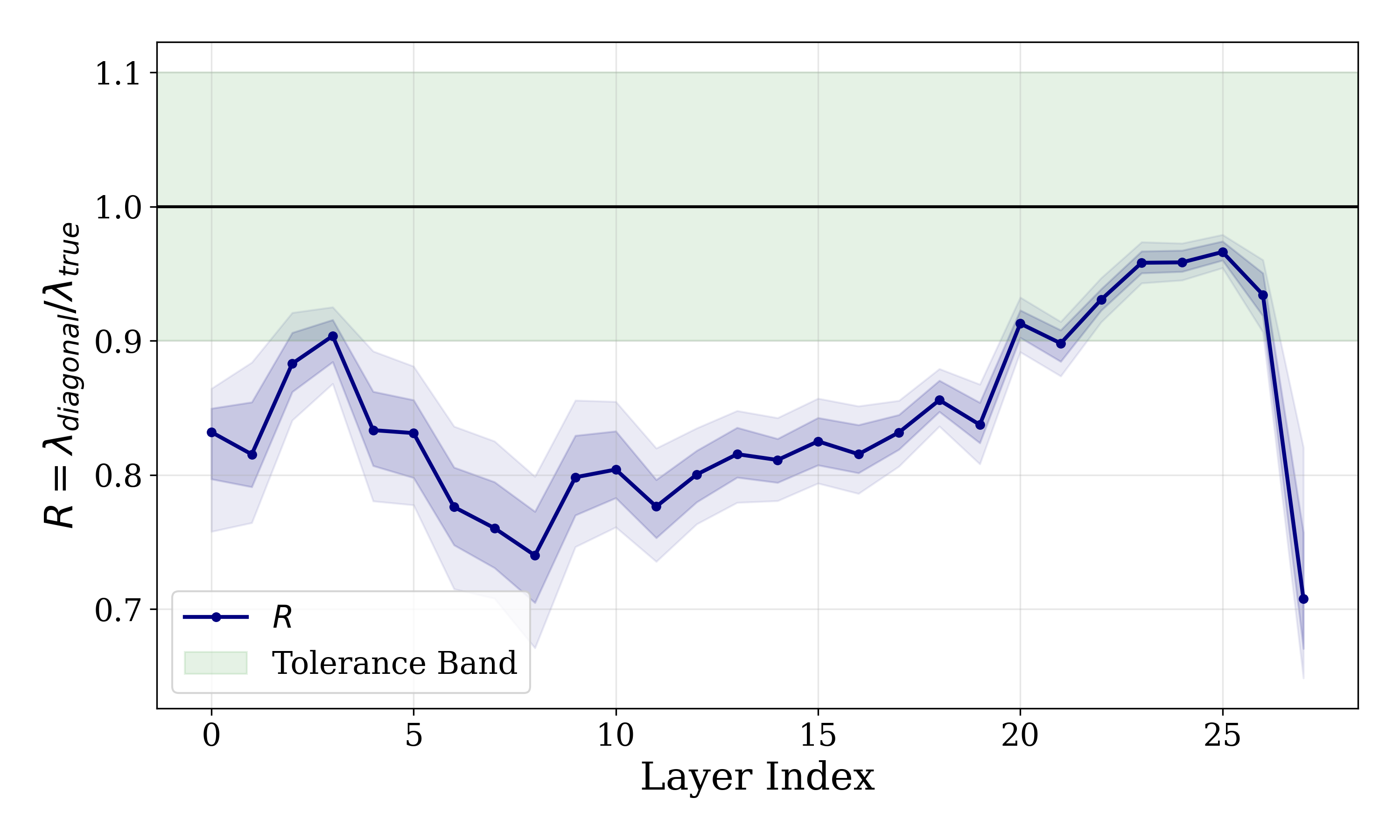}
\caption{Assumption 1: diagonal dominance ($R$)}
\end{subfigure}\hfill
\begin{subfigure}[b]{0.48\linewidth}
\centering
\includegraphics[width=\linewidth]{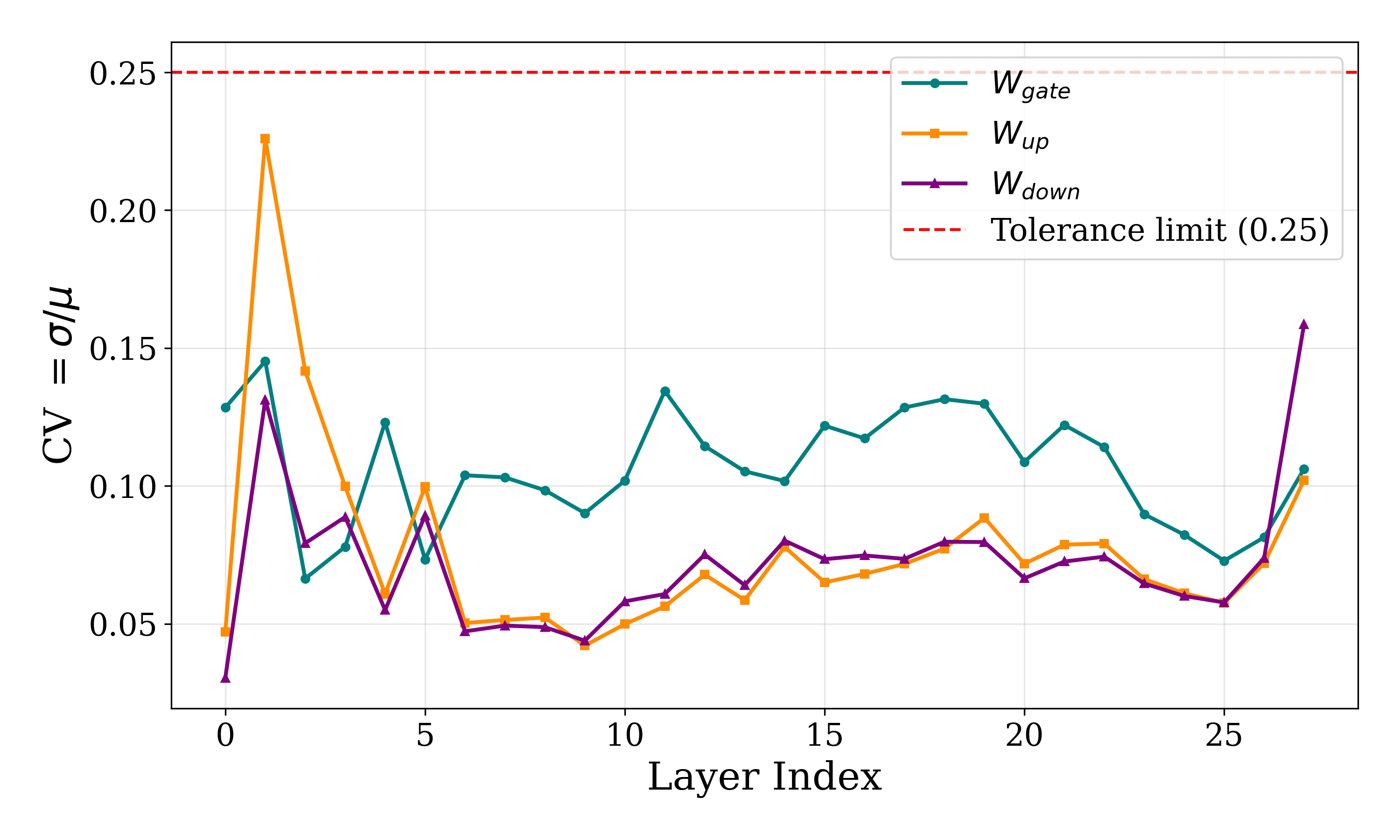}
\caption{Assumption 2: weight-norm uniformity (CV)}
\end{subfigure}
\caption{Numerical validation of the two mean-field assumptions, evaluated layer-by-layer. The tolerance bands mark $R$ close to $1$ and $\text{CV}\ll1$. Shaded bands show the 10th--90th (light) and 25th--75th (dark) percentile ranges.}
\label{fig:mft-num-check}
\end{figure}

Both assumptions hold reasonably well across the network, with the largest deviations appearing in the last few layers (Fig.~\ref{fig:mft-num-check}); the formula is, in any case, ultimately justified by its close empirical agreement with the exact measurement (Fig.~\ref{fig:mft}). A more detailed numerical study of these approximations is given in the underlying thesis~\cite{jaca2026thesis}.

\section{Sliding-Windows}
\label{sec:sliding_appendix}

\begin{figure*}
    \centering
    \begin{subfigure}[b]{0.32\textwidth}
        \centering
        \includegraphics[width=\linewidth]{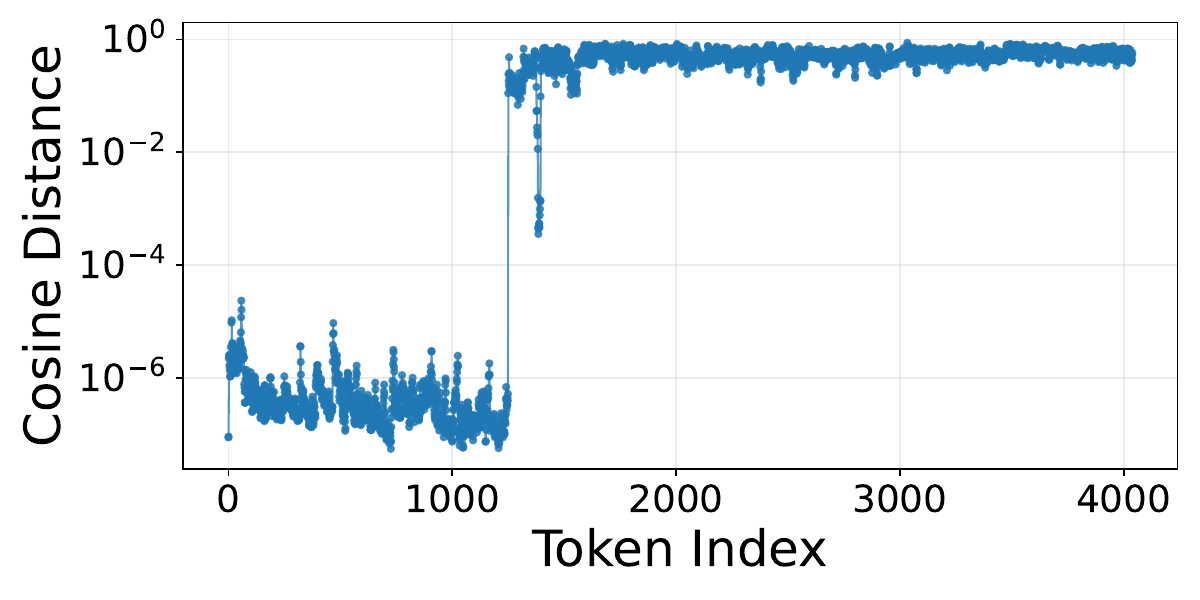}
        \caption{Hidden state, window size 1}
    \end{subfigure}
    \begin{subfigure}[b]{0.32\textwidth}
        \centering
        \includegraphics[width=\linewidth]{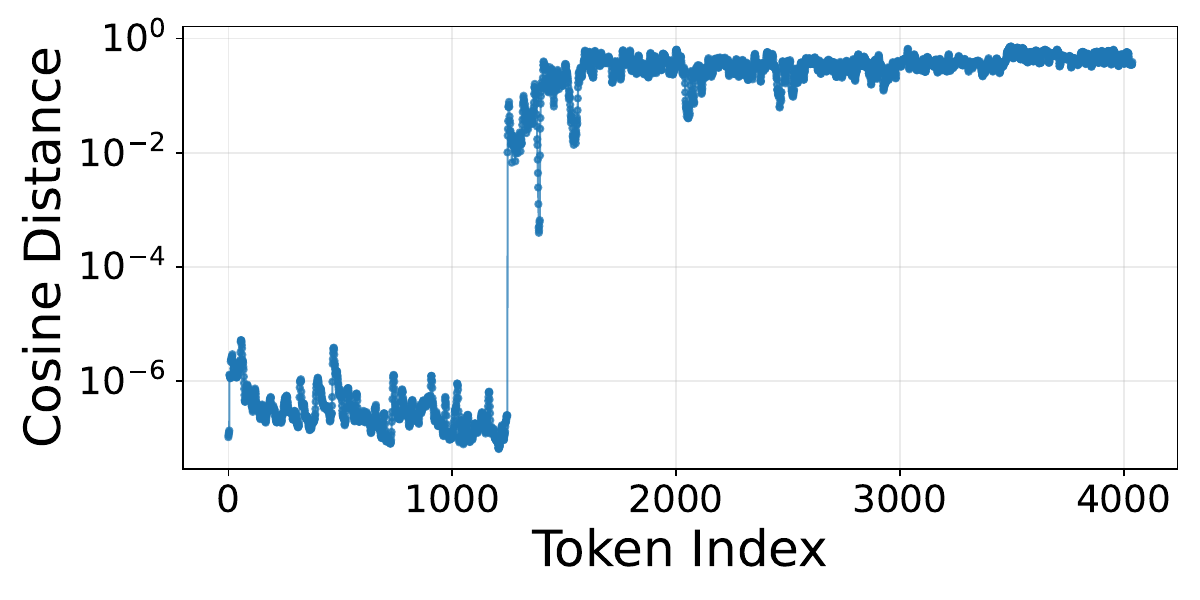}
        \caption{Hidden state, window size 4}
    \end{subfigure}
    \begin{subfigure}[b]{0.32\textwidth}
        \centering
        \includegraphics[width=\linewidth]{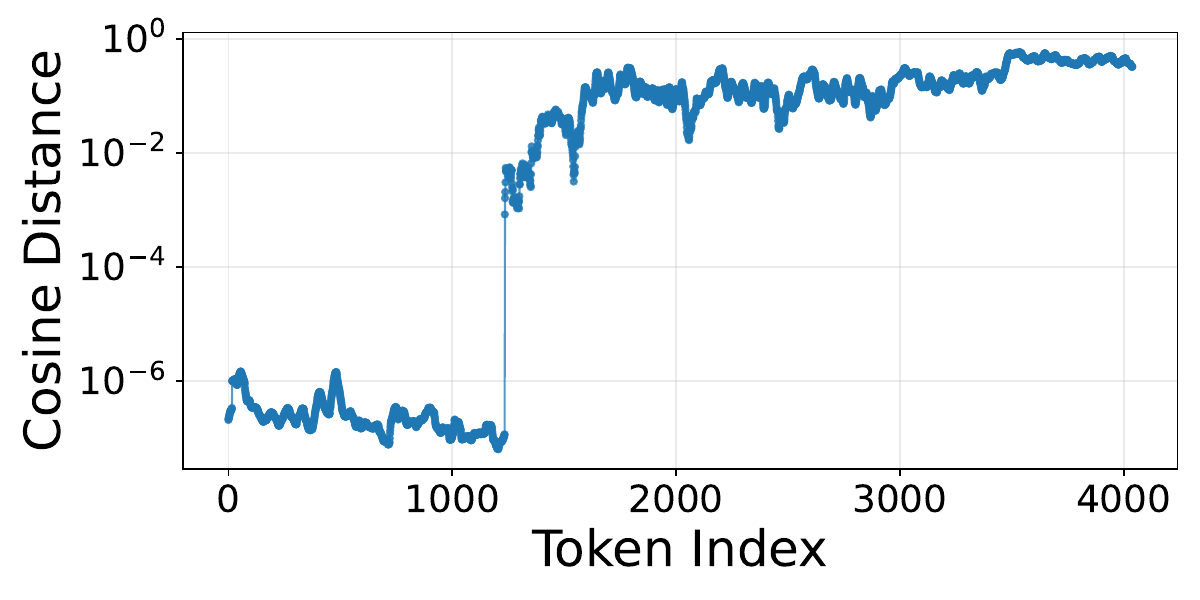}
        \caption{Hidden state, window size 16}
    \end{subfigure}
    \\[0.5em]
    \begin{subfigure}[b]{0.32\textwidth}
        \centering
        \includegraphics[width=\linewidth]{cos_timeseries_0_2_window_size_1.pdf}
        \caption{Embedding, window size 1}
    \end{subfigure}
    \begin{subfigure}[b]{0.32\textwidth}
        \centering
        \includegraphics[width=\linewidth]{cos_timeseries_0_2_window_size_4.pdf}
        \caption{Embedding, window size 4}
    \end{subfigure}
    \begin{subfigure}[b]{0.32\textwidth}
        \centering
        \includegraphics[width=\linewidth]{cos_timeseries_0_2_window_size_16.pdf}
        \caption{Embedding, window size 16}
    \end{subfigure}
    \caption{Divergence between two trajectories using Hausdorff distance, for different sliding window sizes. Top row: hidden state space. Bottom row: sentence embedding space.}
    \label{fig:sliding_window_compare}
\end{figure*}

Figure \ref{fig:sliding_window_compare} illustrates the divergence between two nearby trajectories measured using the Hausdorff distance for different sliding window sizes. In the hidden-state space (top row), the window-free case (window size 1) exhibits strong high-frequency fluctuations, indicating substantial measurement noise. Increasing the window size progressively smooths the divergence curve and reveals a more coherent growth trend. In contrast, trajectories constructed in sentence embedding space (bottom row) display substantially reduced noise even without windowing, with comparatively minor qualitative changes as the window size increases.

This behavior indicates that the hidden state associated with a single token primarily encodes information local to that token, rather than a complete summary of the full contextual history. Although each hidden state is influenced by preceding tokens through self-attention, it does not constitute a sufficient representation of the system’s effective state when considered in isolation. By contrast, sentence embeddings aggregate semantic information over longer spans of text, leading to inherently smoother trajectory representations.

These observations imply that the effective dynamical state of an LLM during inference is distributed across the entire context window rather than being localized in the hidden state of the most recent token alone. Consequently, analyses based solely on instantaneous hidden states correspond to studying a projection of the full system dynamics. Incorporating sliding windows partially reconstructs the underlying state by accounting for temporal dependencies, rendering the dynamics closer to those of a delay-coordinate system. This interpretation is consistent with viewing autoregressive LLM inference as a high-dimensional system with memory, rather than a Markovian process in the space of single-token hidden states.

\section{Dimensionality of the LLM Vocabulary}
\label{sec:appendix_vocab_dim}

As a complementary analysis, we estimate the correlation dimension of the model’s embedding (and unembedding) matrix, which represents the vocabulary as a static point cloud in $\mathbb{R}^D$ rather than a trajectory. This analysis benefits from a large number of points (approximately $1.5\times10^5$) without requiring inference.

The estimated correlation dimension of the vocabulary embedding is substantially lower than that of random vectors in the same ambient dimension and differs from the dimensions observed for hidden-state trajectories. This suggests that the learned vocabulary occupies a structured, low-dimensional manifold within the embedding space. However, trajectories generated by the model do not correspond to simple random walks on this manifold; instead, they are produced by nonlinear transformations that generate novel latent states not present in the vocabulary itself.

\begin{figure}[H]
    \centering
    \includegraphics[width=0.8\linewidth]{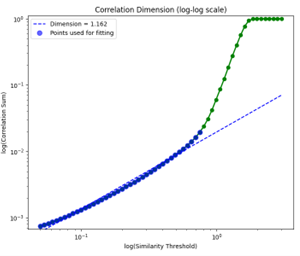}
    \caption{Correlation dimension of the LLM's embedding matrix.}
    \label{fig:embedding_dim}
\end{figure}

\section{Experiments' details}
\label{sec:experiments}

For this study, we used the following experimental setup:
\begin{itemize}
    \item \textbf{Model:} deepseek-ai/DeepSeek-R1-Distill-Qwen-1.5B \cite{deepseek2025r1distillqwen1.5b} \footnote{The model used is distilled from DeepSeek R1 reducing the number of parameters from 671 billion to 1.5 billion. It is trained and fine-tuned to replicate the behavior from the DeepSeek R1. This distilled model is based on the Qwen2.5 family and there are differences in the architecture, which are beyond the scope of this study. All the points mentioned in Section \ref{subsec:llms_dynamics} apply to both, and generally to most LLMs. The reason we used this model was memory limitations in the GPUs available for this study.}
    \item \textbf{Temperature:} 0 (to ensure determinism)
    \item \textbf{Sampling:} greedy (to ensure determinism)
    \item \textbf{Context Window:} 3096 tokens
    \item \textbf{Initial Perturbation Magnitude:} 0.00035 or 0.0004 
    \item \textbf{Vocabulary Size:} 151936
    \item \textbf{Hidden/Embedding Dimension:} 1536
    \item \textbf{Attention Heads:} 12
    \item \textbf{Hidden Layers:} 28
\end{itemize}

The choice of the initial perturbation magnitude $r$ is not arbitrary: its admissible range is bounded on both sides by the dynamics itself. From below, the perturbation must exceed the FP16 numerical noise floor of the inference pipeline ($\approx 5\times10^{-4}$ in activation space, cf.\ Sec.~\ref{subsec:layer_perturbations}), below which it is indistinguishable from rounding noise. From above, a perturbation large enough to flip the very first sampled token destroys the latent-divergence regime under study, reducing the experiment to comparing two unrelated prompts. The chosen values lie in the intermediate window, where the layer-wise response is still linear in $r$ (Fig.~\ref{fig:scaling}, fitted exponent $\approx1$), so the reported phenomenology does not depend on the precise value of $r$: it plays the role of the infinitesimal perturbation of classical Lyapunov analysis, made as small as the hardware permits.

\section{Example prompts}
\label{sec:ex_prompt}

\begin{promptblock}
Provide a comprehensive technical review of current and proposed propulsion systems for interstellar travel. Compare chemical rockets, nuclear propulsion, laser sails, antimatter drives, and other theoretical concepts in terms of energy requirements, achievable speeds, technological feasibility, and projected timelines for development. Include discussion of major projects in the history of the field.
\end{promptblock}

\begin{promptblock}
Examine how childhood experiences shape personality development. Discuss various influences including family environment, education, friendships, and significant life events. Explain psychological concepts like attachment theory and nature vs. nurture in accessible terms. Provide examples of how positive and negative experiences can affect adult personality traits and behaviors.
\end{promptblock}

\end{document}